\documentclass{aa}
\usepackage[pdftex]{graphicx}
\usepackage[flushleft]{threeparttable}
\usepackage{amsmath}
\usepackage{hyperref}
 \usepackage{color}
\usepackage{subfigure}
\newcommand{\HI}{\mbox{H{\sc i}}}
\usepackage{color}

\begin{document}      

   \title{The uneven sisters I}
   \subtitle{NGC~4388 -- a strongly constrained ram pressure stripping event}

   \author{B.~Vollmer\inst{1}, C.~Pappalardo\inst{2,3}, M. Soida\inst{4}, \& A.~Lan\c{c}on\inst{1}}

   \offprints{B.~Vollmer, e-mail: Bernd.Vollmer@astro.unistra.fr}

   \institute{Universit\'e de Strasbourg, CNRS, Observatoire astronomique de Strasbourg, UMR 7550, F-67000 Strasbourg, France \and 
     Centro de Astronomia e Astrof\'{i}sica da Universidade de Lisboa, Observat\'{o}rio Astron\'{o}mico de Lisboa, Tapada da Ajuda, 1349-018 Lisboa, Portugal
     \and Instituto de Astrof\'{i}sica e Ciencias do Espa\c{c}o, Universidade de Lisboa, OAL, Tapada da Ajuda, 1349-018 Lisboa, Portugal
     \and Astronomical Observatory, Jagiellonian University, 30-244, Krak\'{o}w, Poland} 

   \date{Received / Accepted}

   \authorrunning{Vollmer et al.}
   \titlerunning{The uneven sisters I: NGC~4388}

\abstract{
Since the Virgo cluster is the closest galaxy cluster in the northern hemisphere, galaxy interactions can be observed in it with 
a kpc resolution. 
The spiral galaxy NGC~4388 underwent a ram pressure stripping event $\sim 200$~Myr ago caused by its highly eccentric orbit within the Virgo cluster.
This galaxy fulfills all diagnostic criteria for having undergone active ram pressure stripping in the recent past:
a strongly truncated H{\sc i} and H$\alpha$ disk, an asymmetric ridge of polarized radio continuum emission,
extended extraplanar gas toward the opposite side of the ridge of polarized radio continuum emission, and a
recent (a few $100$~Myr) quenching of the star formation activity in the outer, gas-free galactic disk.
We made dynamical simulations of the ram pressure stripping event to investigate the influence of galactic structure on the
observed properties of NGC~4388. 
The combination of a deep optical spectrum of the outer gas-free region of the galactic disk together with deep H{\sc i},
H$\alpha$, FUV, and polarized radio continuum data permits to constrain numerical simulations to derive the temporal ram pressure profile, the 3D
velocity vector of the galaxy, and the time since peak ram pressure with a high level of confidence.
From the simulations an angle between the ram pressure wind and the galactic disk of $30^{\circ}$ is derived.
The galaxy underwent peak ram pressure  $\sim 240$~Myr ago.
The observed asymmetries in the disk of NGC~4388 are not caused by
the present action of ram pressure, but by the resettling of gas that has been pushed out of the galactic disk during the ram pressure stripping event.
For the detailed reproduction of multi-wavelength observations of a spiral galaxy that undergoes or underwent a ram pressure stripping event,
galactic structure, i.e. spiral arms, has to be taken into account.
}

\maketitle

\section{Introduction \label{sec:intro}}

The environment has a strong influence on spiral galaxies evolving in a cluster of galaxies.
The ideal place to study the evolution of cluster spiral galaxies is the Virgo cluster,
because it represents the only cluster in the northern hemisphere where one can observe the ISM distribution and 
kinematics of cluster galaxies at a kpc resolution (1~kpc $\sim 12''$)\footnote{We assume a distance of 17~Mpc to the Virgo cluster}.
The Virgo cluster is dynamically young and spiral-rich.
The cluster spiral galaxies have lost up to $90$\,\% of their neutral hydrogen, i.e. they are
H{\sc i} deficient (Chamaraux et al. 1980, Giovanelli \& Haynes 1983).
Imaging H{\sc i} observations have shown that these galaxies have truncated H{\sc i} disks 
(Cayatte et al. 1990, Chung et al. 2009). Thus, the cluster environment changes the H{\sc i} content and morphology of 
Virgo cluster spiral galaxies.

Vollmer et al. (2007) showed that the distribution of polarized radio continuum emission of cluster spiral galaxies
can be very asymmetric compared to the emission of field galaxies. Asymmetric ridges of polarized radio emission are produced
by compression or shear motions (Otmianowska-Mazur \& Vollmer 2003). Both kinds of motions occur during ram pressure stripping events
or tidal interactions. In the case of a ram pressure stripping event the gas is compressed on one side of the galactic disk
and pushed out of the disk on the other side, as observed in NGC~4522 (Kenney et al. 2004). Together with the gas distribution and
velocity field, the polarized radio continuum emission thus represents an important diagnostic tool for ram pressure stripping (e.g., Vollmer et al. 2006).

Based on these diagnostic tools Vollmer (2009) was able to establish a time sequence for ram pressure stripping in the Virgo cluster:
NGC~4501 is observed before peak ram pressure, NGC~4522 and NGC~4438 close to peak ram pressure, NGC~4388 $\sim 100$~Myr after peak ram pressure,
and NGC~4569 $\sim 300$~Myr after peak ram pressure.

Two edge-on spiral galaxies, NGC~4388 and NGC~4402, are located close ($\sim 400$~kpc) to the cluster center (M~87) and close ($135$~kpc) to each other in projection on the sky.
Both galaxies have high radial velocities with respect to the cluster mean velocity ($v_{\rm N4388} \sim 1500$~km\,s$^{-1}$, $v_{\rm N4402} \sim -900$~km\,s$^{-1}$).
Both galaxies are strongly H{\sc i} deficient, their disks of atomic hydrogen being strongly truncated. Whereas NGC~4402 shows only a small
H{\sc i} tail (Chung et al. 2009), NGC~4388 has a spectacular H{\sc i} plume extending $\sim 100$~kpc to the north of the galaxy (Oosterloo \& van Gorkom 2005).
The part of the plume close to the galactic disk is also visible in H$\alpha$ emission (Yoshida et al. 2002). The polarized radio continuum emission distributions of
both galaxies are asymmetric with ridges of polarized emission located at the southern edges of the galaxies (Vollmer et al. 2007). 
In NGC~4388 the distribution of polarized radio continuum emission is complex due to a nuclear outflow. 

Both galaxies thus share common key properties: asymmetric ridges of polarized radio continuum emission and strongly truncated H{\sc i} disks
together with extraplanar neutral gas.
The huge H{\sc i} tail of NGC~4388 indicates that the galaxy is stripped for the first time, i.e. it was not significantly H{\sc i} deficient before
its passage in the Virgo cluster core. Does NGC~4402 share the same history in the Virgo cluster? Is it also stripped for the first time?
Within this article on NGC~4388 and its companion on NGC~4402, we want to find answers to these questions.

In order to verify the time to peak ram pressure of NGC~4388 derived from the direct comparison of the model gas distribution, 
Pappalardo et al. (2010) took a deep optical spectrum of the outer gas-free stellar disk of NGC~4388.
With the help of stellar population synthesis models they could determine the time since star formation was quenched in the stellar disk 
when the gas was removed by ram pressure stripping. This time ($\sim 200$~Myr) was in agreement with the star formation quenching time
derived from Lick indices by Crowl et al. (2008), but in disagreement with the time to peak ram pressure
determined by Vollmer \& Huchtmeier (2003) based on Effelsberg H{\sc i} data. 

In this article we present new dynamical simulations of the ISM of NGC~4388 together with calculations of the associated magnetic field and
the polarized radio continuum emission. We study the influence of galactic structure, i.e. spiral arms, on the final gas distribution and velocity
field, the H$\alpha$, FUV, and polarized radio continuum emission, and the optical spectrum after the ram pressure stripping event.
The comparison of the simulated data with observations can discriminate between the different model scenarios.

\section{The model \label{sec:model}}

We use the N-body code described in Vollmer et al. (2001) which consists of 
two components: a non-collisional component
that simulates the stellar bulge/disk and the dark halo, and a
collisional component that simulates the ISM.
A new scheme for star formation has been implemented, where stars are formed
during cloud collisions and are then evolved as non-collisional particles.

\subsection{Halo, stars, and gas} 

The non--collisional component consists of 81\,920 particles, which simulate
the galactic halo, bulge, and disk.
The characteristics of the different galactic components are shown in
Table~\ref{tab:param}.
\begin{table}
      \caption{Total mass, number of particles $N$, particle mass $M$, and smoothing
        length $l$ for the different galactic components.}
         \label{tab:param}
      \[
         \begin{array}{lllll}
           \hline
           \noalign{\smallskip}
           {\rm component} & M_{\rm tot}\ ({\rm M}$$_{\odot}$$)& N & M\ ({\rm M}$$_{\odot}$$) & l\ ({\rm pc}) \\
           \hline
           {\rm halo} & 1.9 \times 10$$^{11}$$ & 32768 & $$5.9 \times 10^{6}$$ & 1200 \\
           {\rm bulge} & 6.7 \times 10$$^{9}$$ & 16384 & $$4.1 \times 10^{5}$$ & 180 \\
           {\rm disk} & 3.3 \times 10$$^{10}$$ & 32768 & $$1.0 \times 10^{6}$$ & 240 \\
           \noalign{\smallskip}
        \hline
        \end{array}
      \]
\end{table}
The resulting rotation velocity is $v_{\rm rot} = 190$~km\,s$^{-1}$ and the rotation curve
becomes flat at a radius of about $6$~kpc. 

We have adopted a model where the ISM is simulated as a collisional component,
i.e. as discrete particles or clouds which possess a mass and a radius and which
can have inelastic collisions (sticky particles).
Since the ISM is a turbulent and fractal medium (see e.g. Elmegreen \& Falgarone 1996),
it is neither continuous nor discrete. The volume filling factor of the warm and cold phases
is smaller than one. The warm neutral and ionized gas fill about $30-50\%$ of the volume,
whereas cold neutral gas has a volume filling factor smaller than 10\% (Boulares \& Cox 1990). 
It is not clear how this fraction changes, when an external 
pressure is applied. In contrast to smoothed particles hydrodynamics (SPH), which is a 
quasi continuous approach and where the particles cannot penetrate each other, our approach 
allows a finite penetration length, which is given by the mass-radius relation of the particles.
Both methods have their advantages and their limits.
The advantage of our approach is that ram pressure can be included easily as an additional
acceleration on particles that are not protected by other particles (see Vollmer et al. 2001).

The 20\,000 particles of the collisional component represent gas cloud complexes which are 
evolving in the gravitational potential of the galaxy.
The total assumed gas mass is $M_{\rm gas}^{\rm tot}=4.4 \times 10^{9}$~M$_{\odot}$,
which corresponds to the total neutral gas mass before stripping, i.e.
to an \HI\ deficiency of 0, which is defined as the logarithm of the ratio between
the \HI\ content of a field galaxy of same morphological type and diameter
and the observed \HI\ mass (e.g., Chung et al. 2009).
To each particle a radius is assigned depending on its mass. 
During the disk evolution the particles can have inelastic collisions, 
the outcome of which (coalescence, mass exchange, or fragmentation) 
is simplified following Wiegel (1994). 
This results in an effective gas viscosity in the disk. 

The optical radius of NGC~4388 is $R_{25}=2.42' \sim 12$~kpc.
If we assume an inclination-corrected optical radius of $10$--$11$~kpc, the initial radial model H{\sc i} profile with respect to $R/R_{25}$ is similar to that of 
the local field spiral galaxy NGC~5055 (Leroy et al. 2008). It can be broadly described by an exponential with a lengthscale of $5.5$~kpc
with two dips at $2.5$~kpc and $7.5$~kpc. With a molecular fraction proportional to the square root of the gas density $f_{\rm mol} \propto \sqrt{\rho}$
(Vollmer et al. 2008), the H{\sc i} surface density profile is constant from $2.5$~kpc to $7.5$~kpc ($\Sigma_{\rm HI} \sim 10$~M$_{\odot}$pc$^{-2}$),
declines rapidly to $\Sigma_{\rm HI} \sim 4$~M$_{\odot}$pc$^{-2}$ and then stays approximately constant between $9$~kpc and $14$~kpc.

As the galaxy moves through the ICM, its clouds are accelerated by
ram pressure. Within the galaxy's inertial system its clouds
are exposed to a wind coming from a direction opposite to that of the galaxy's 
motion through the ICM. 
The adopted temporal ram pressure profile has the form of a Lorentzian,
which is realistic for galaxies on highly eccentric orbits within the
Virgo cluster (Vollmer et al. 2001).
The effect of ram pressure on the clouds is simulated by an additional
force on the clouds in the wind direction. Only clouds which
are not protected by other clouds against the wind are affected.
Since the gas cannot develop instabilities, the influence of turbulence 
on the stripped gas is not included in the model. The mixing of the
intracluster medium into the ISM is very crudely approximated by a finite
penetration length of the intracluster medium into the ISM, i.e. until this penetration
length the clouds undergo an additional acceleration due to ram pressure.

We note that the treatment of the interstellar medium in numerical simulations is very crude (gas clouds
having partially inelastic collisions). As soon as the clouds are stripped from the galaxies, the clouds mainly behave like ballistic
particles which are affected by ram pressure. For the acceleration by ram pressure, we assume $a=p_{\rm ram}/\Sigma$,
where the surface density of the atomic gas clouds is assumed to be $\Sigma=10$~M$_{\odot}$pc$^{-2}$ at all times.
Cloud evaporation by or mixing with the hot intracluster medium are thus not included in our simulations.
Because this kind of clouds are expected to be stripped more efficiently, a larger extent of the outer H{\sc i}
tail than that calculated with clouds of constant surface density might be expected.

The particle trajectories are integrated using an adaptive timestep for
each particle. This method is described in Springel et al. (2001).
The following criterion for an individual timestep is applied:
\begin{equation}
\Delta t_{\rm i} = \frac{20~{\rm km\,s}^{-1}}{a_{\rm i}}\ ,
\end{equation}
where $a_{i}$ is the acceleration of the particle i.
The minimum value of $t_{\rm i}$ defines the global timestep used 
for the Burlisch--Stoer integrator that integrates the collisional
component. It is typically a few $10^{4}$~yr.

\subsection{Star formation \label{sec:sfr}}

We assume that the star formation rate is proportional to the cloud collision rate.
During the simulations stars are formed in cloud-cloud collisions. At each collision
a collisionless particle is created which is added to the ensemble of collisional and
collisionless particles. The newly created collisionless particles have zero mass
(they are test particles) and the positions and velocities of the colliding clouds after the collision. 
These particles are then evolved passively with the whole system. 
Since in our sticky-particle scheme there is mass exchange, coalescence, or fragmentation at the
end of a cloud--cloud collision, the same clouds will not collide infinitely.
The local collision rate traces the cloud density and the velocity dispersion of the collisional component.
Since the cloud density increases rapidly with decreasing galactic radius, the number of newly created particles 
rises steeply toward the galaxy center.

The information about the time of creation is attached to each newly created star particle.
In this way the H$\alpha$ emission distribution can be modeled by the distribution of
star particles with ages less than $10$~Myr. The UV emission of a star particle in the two GALEX bands
is modeled by the UV flux from single stellar population models from STARBURST99 (Leitherer et al. 1999).
We assume that all star particles have the number of UV-emitting stars.
The total UV distribution is then the extinction-free distribution of the UV emission of the newly created
star particles. For the unperturbed galaxy the resulting power law between star formation based on the model UV emission 
and the total/molecular gas surface density has an exponent of $1.7$/$1.2$, respectively
(Fig.~A.1 of Vollmer et al. 2012). This is close to the observational findings of Bigiel et al. (2008).

\subsection{The influence of the galaxy orbit on the simulations \label{sec:orbits}}

\begin{figure*}
  \resizebox{\hsize}{!}{\includegraphics{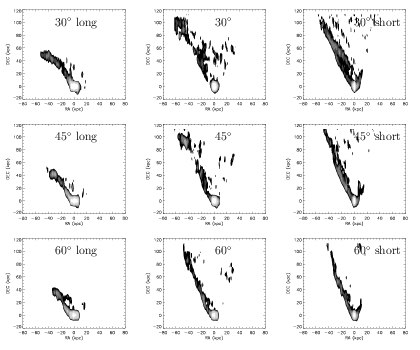}}
        \caption{Large-scale H{\sc i} distribution of models with different cluster orbits and wind angles ($30^{\circ}$, $45^{\circ}$, $60^{\circ}$) 
          the angle between the ram pressure wind and the galactic disk. The projection angles are $i=90^{\circ}$ (edge-on) and $PA=90^{\circ}$ (horizontal).
          The Lorentzian ram pressure profiles with shorter widths have higher values of peak ram pressure. Contour levels are
          $(3,6,9,15,30,150) \times 10^{19}~{\rm cm}^{-2}$. The times of interest are $240$~Myr, $210$~Myr, and $200$~Myr after peak ram pressure for wind angles of
          $30^{\circ}$, $45^{\circ}$, and $60^{\circ}$, respectively.
        } \label{fig:zusammen}
\end{figure*}

As a first test we investigated the influence of the galaxy orbit within the Virgo cluster on the formation
of the stripped H{\sc i} tail. To do so, we adopted three different galaxy orbits from Vollmer et al. (2001; Fig.~2).
These orbits lead to Lorentzian ram pressure profiles (Vollmer et al. 2001; Fig~3), which are characterized by the
ram pressure maximum $p_{\rm max}$ and the width $\Delta t$. The three different orbits have long, intermediate,
and short timescales: $(p_{\rm max},\Delta t)=(1250,200),\ (5000, 100),\ (2 \times 10^4, 50)$, where $p_{\rm max}$ and $\Delta t$ are
in cm$^{-3}$/(km\,s$^{-1}$)$^2$ and Myr, respectively. In addition, we varied the wind angle, the angle between the
ram pressure wind and the disk plane, between $30^{\circ}$ and $60^{\circ}$.
H{\sc i} column density maps were produced $200$, $210$, and $240$~Myr after the time of peak ram pressure for the models
with angles between the galactic disk and the ram pressure wind of $60^{\circ}$, $45^{\circ}$, and $30^{\circ}$, respectively (Fig.~\ref{fig:zusammen}).

Since the peak ram pressure sets the gas truncation radius, the H{\sc i} disks of the $(p_{\rm max},\Delta t)=(1250,200),\ (2 \times 10^4, 50)$
simulations are too large and too small compared to the VIVA H{\sc i} observations (Chung et al. 2009).
Moreover, the model H{\sc i} tails of the low peak ram pressure simulations are about two times smaller than the 
observed H{\sc i} tail (Oosterloo \& van Gorkom 2005). On the other hand, the model H{\sc i} tails of the 
high peak ram pressure simulations are too long and too thin compared to observations. The high peak ram pressure simulations also
show a small tail on the western side of the galactic disk, which is not observed.

Only the simulations with an intermediate peak ram pressure and an intermediate width show an H{\sc i} tail length which is
consistent with observations. A wind angle of $60^{\circ}$ leads to a tail width which is significantly smaller than that of the
observed H{\sc i} tail. The observed extent in east-west direction is only approximately reproduced by the simulations with wind
angles of $30^{\circ}$ and $45^{\circ}$. These models show faint emission west of the tail which is not observed.
Moreover, all models miss the observed gas at a height of $\sim 80$~kpc, $\sim -35$~kpc east of the galaxy center.

We conclude that the simulations which are most consistent with H{\sc i} observations have $p_{\rm max} \sim 8 \times 10^{-11}$~dyn\,cm$^{-2}$,
a width $\Delta t = 100$~Myr, and a wind angle of $30^{\circ}$ to $45^{\circ}$. 
The ram pressure profile of the model of Vollmer \& Huchtmeier (2003) had the same peak ram pressure, but a two times smaller width and a wind angle of $45^{\circ}$. 

\subsection{The influence of galactic structure on the simulations}

\begin{figure}
  \resizebox{\hsize}{!}{\includegraphics{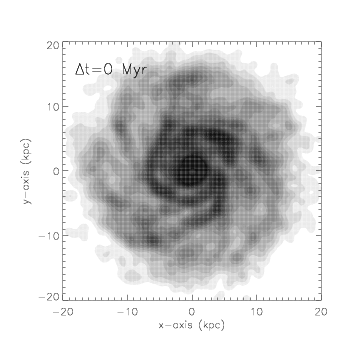}\includegraphics{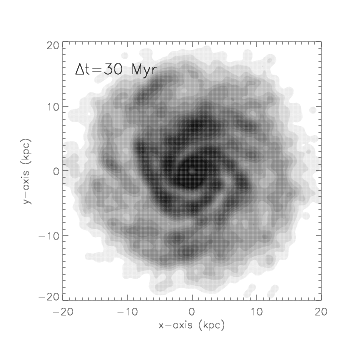}}
  \resizebox{\hsize}{!}{\includegraphics{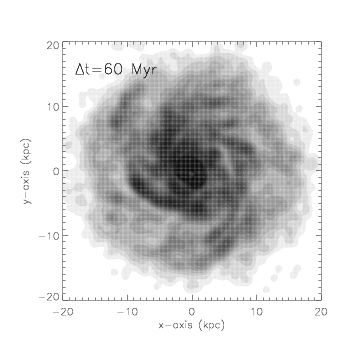}\includegraphics{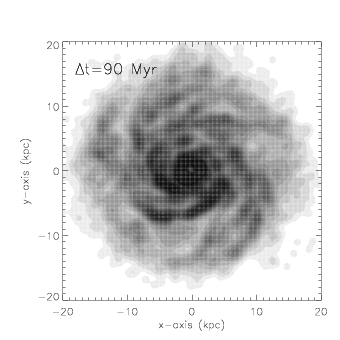}}
        \caption{Snapshots of a galaxy after 0, 30, 60, and 90~Myr of evolution with a wind angle of $30^{\circ}$, 590, 560, 530, and 500~Myr before peak ram pressure.
        } \label{fig:foviews}
\end{figure}

As in Nehlig et al. (2016), we study the influence of galactic structure on the outcome of a ram pressure stripping event
by using the same initial condition and fixing the peak of ram pressure stripping at different times: $t_{\rm peak}=500,\ 510,\ 530,\ 550,\ 570,\ 590$~Myr
after the beginning of the simulation (see Table~\ref{tab:results}).
For comparison, the time interval of $100$~Myr corresponds to approximately half a rotation period at the distance of the main spiral arms ($\sim 7$~kpc).
The time of interest is $210$~Myr after peak ram pressure for the wind angle of $45^{\circ}$.
For the simulations with a wind angle of $30^{\circ}$ the timesteps of peak ram pressure are $t_{\rm peak}=500,\ 530,\ 560,\ 590$~Myr after the beginning 
of the simulation. The time of interest is $240$~Myr after peak ram pressure.
These timesteps were chosen to reproduce the large-scale H{\sc i} distribution of NGC~4388. Table~\ref{tab:results} gives the numbering of the different models
(6, 5, 1, 2, 3, 4, 7, 8, 10 for increasing time delays). The face-on projections of the gas disk $500$~Myr before peak ram pressure for 
$t_{\rm peak} = 500,\ 530,\ 560,\ 590$~Myr are presented in Fig.~\ref{fig:foviews}. 

\section{Comparison with observations \label{sec:comparison}}

For comparison with observations, H{\sc i}, H$\alpha$, FUV, and radio continuum maps together with H{\sc i} position-velocity diagrams along the
galaxy's major axis and optical spectra of the stripped gas-free part of the galactic disk were produced from the simulations.

\subsection{Atomic hydrogen}

To separate the atomic and molecular gas phases, we follow Vollmer et al. (2008)  by assuming
that the molecular fraction depends linearly on the square root of the gas density ($f_{\rm mol}=\sqrt{\rho/(0.5\,{\rm M_{\odot}pc^{-3}})}$),
where $\rho$ is the total gas density. Moreover, the molecular fraction cannot exceed unity. The H{\sc i} column density is then given
by $\Sigma_{\rm HI}=(1-f_{\rm mol})\,\Sigma$, where $\Sigma$ is the total gas surface density.

\subsubsection{The H{\sc i} tail}

The observed large-scale H{\sc i} distribution together with the position-velocity diagram of NGC~4388 are presented in
the upper panels of Fig.~\ref{fig:hicomp}. The extended plume has a total extent of $\sim 20' = 100$~kpc. The region of highest
gas surface density is located $\sim 13.5' = 67$~kpc north of the galactic disk and $\sim 7.2' = 36$~kpc east of the center of the galaxy.
The plume region near the galactic disk is almost devoid of detectable H{\sc i} emission.
The velocity range of the plume is roughly $-500$~km\,s$^{-1}$ to $0$~km\,s$^{-1}$ with respect to the galaxy's systemic velocity of
$v_{\rm sys}=2524$~km\,s$^{-1}$. The velocity width is about $100$~km\,s$^{-1}$, except in the region near the maximum where the velocity width is $\sim 200$~km\,s$^{-1}$.  

\begin{figure*}
        \resizebox{\hsize}{!}{\includegraphics{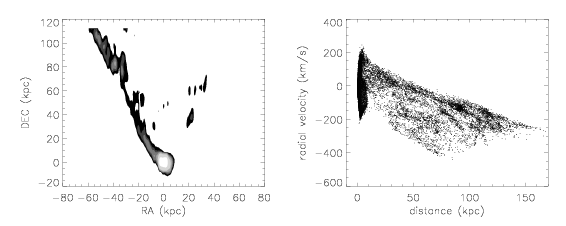}\includegraphics{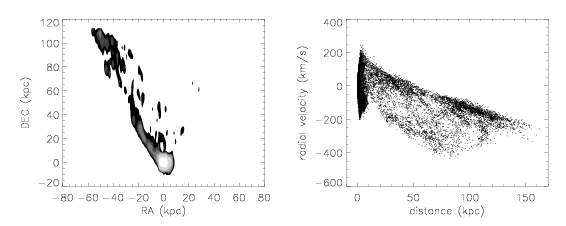}
        \put (-500,50) {\Large model 2}\put (-1060,50) {\Large model 1}}
        \resizebox{\hsize}{!}{\includegraphics{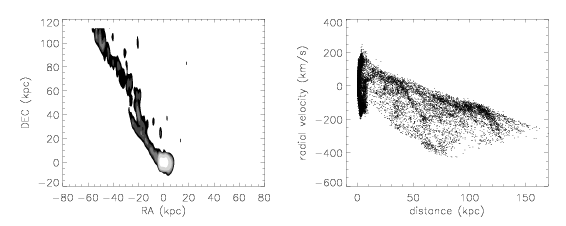}\includegraphics{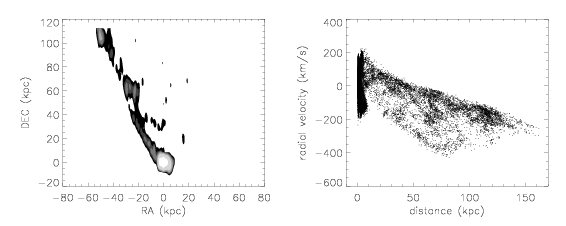}
        \put (-500,50) {\Large model 4}\put (-1060,50) {\Large model 3}}
        \resizebox{\hsize}{!}{\includegraphics{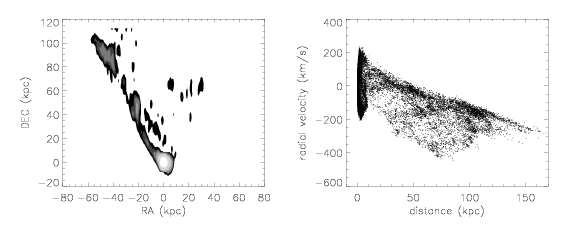}\includegraphics{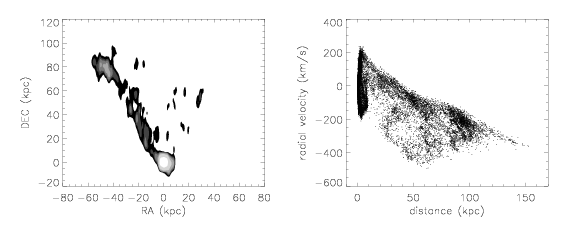}
        \put (-500,50) {\Large model 6}\put (-1060,50) {\Large model 5}}
        \resizebox{\hsize}{!}{\includegraphics{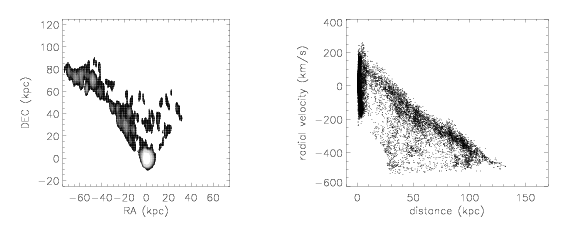}\includegraphics{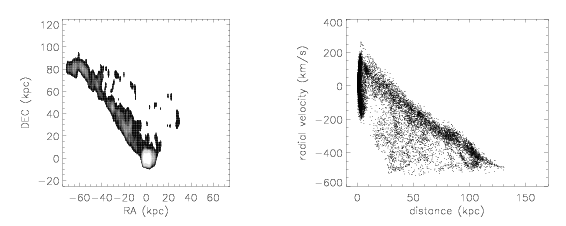}
        \put (-500,50) {\Large model 8}\put (-1060,50) {\Large model 7}}
        \resizebox{\hsize}{!}{\includegraphics{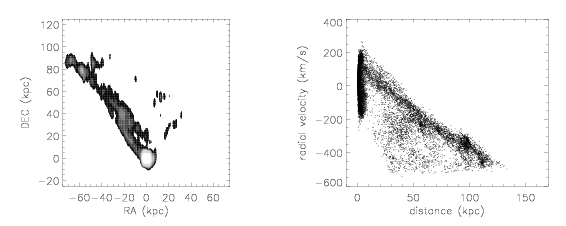}\includegraphics{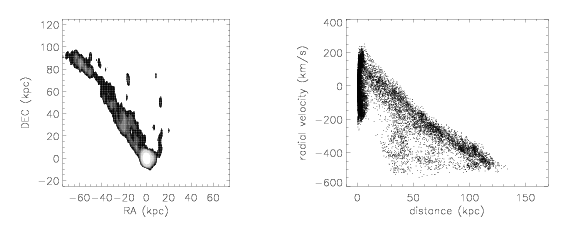}
        \put (-500,50) {\Large model 10}\put (-1060,50) {\Large model 9}}
        \caption{Model H{\sc i} large-scale distributions and position--velocity diagrams with $i=90^{\circ}$. Contour levels are
          $(3,6,9,15,30,150) \times 10^{19}~{\rm cm}^{-2}$.
        } \label{fig:N4388_final_test3_HI1a}
\end{figure*} 

The large-scale H{\sc i} distributions and pV diagrams with respect to the projected distance from the galaxy center 
of all models with wind angles of $45^{\circ}$ and $30^{\circ}$ are presented in Fig.~\ref{fig:N4388_final_test3_HI1a}.
We used an inclination angle of $i=90^{\circ}$ (edge-on) for all model snapshots. The projection with $i=75^{\circ}$ is shown in Fig.~\ref{fig:N4388_final_test3_HI1b}. 
We chose $i=90^{\circ}$, because for this inclination all H{\sc i} tails show a vertical extent between $90$ and $100$~kpc which is close to the observed extent.
On the other hand, for $i=75^{\circ}$ the gas tails are somewhat too extended ($\sim 120$~kpc).
The characteristic curvature of the observed tail at high latitudes is better reproduced by the simulations with a wind angle of $30^{\circ}$.
Models 5, 6, and 7 show low surface density H{\sc i} west of the plume which has no counterpart in observations. The dearth of H{\sc i} emission close
to the galactic disk is reproduced by models 7 and 8. The column density of the model gas tail close to the disk is small for model 9, but higher than observed.
As already stated in Sect.~\ref{sec:orbits}, all models miss the observed gas in the H{\sc i} tail at a height of $\sim 70$~kpc, $\sim -35$~kpc east of the galaxy center.
Only the simulations with a wind angle of $30^{\circ}$ reproduce the observed velocity range of the H{\sc i} plume ($\Delta v \sim 500$~km\,s$^{-1}$).
We thus conclude that the models with a wind angle of $30^{\circ}$ are in better agreement with the data than those with a wind angle of $45^{\circ}$.

For the sake of illustration and clarity, we chose for each observable a ``best-fit'' model (model 9 or 10) and compare it to observations.
The ``best-fit'' models thus can vary for different observables, since we cannot expect that one single model produces all observables given
the degeneracies between galactic structure, time of peak ram pressure, wind angle, and temporal ram pressure profile.
We also use two different inclination angles, $i=75^{\circ}$ and $i=90^{\circ}$. The set of model snapshots for a given inclination, which includes the ``best-fit'' model, 
is shown in the main part of the article, the other set is presented in the Appendix.

The ``best-fit'' model large-scale H{\sc i} distributions together with the position-velocity diagram of NGC~4388 are presented in
the lower panels of Fig.~\ref{fig:hicomp} for an inclination angle of $i=90^{\circ}$. 
Since we restricted ourselves to model~9 and 10 for the ``best-fit'' model, we present model~9 as the ``best-fit'' model.
The model gas plume shows about the same extent and the same east-west asymmetry as the observed gas plume.
The regions of highest surface density in the model gas tail are located at vertical distances of $\sim 80$~kpc and $\sim 40$~kpc,
$55$~kpc and $25$~kpc east of the galaxy center (positions (-55~kpc, 80~kpc) and (-25~kpc, 40~kpc)). 
Whereas the high surface density gas at a vertical distance of $40$~kpc has an observed
counterpart, the observed gas surface density in the region at vertical distance of $80$~kpc is much lower than that of the model.
Whereas the observed tail does not show significant emission close to the galactic disk ($z < 20$~kpc), there is gas of relatively high
surface density in this area of the model tail. Moreover, the low surface density gas in the model tail above the galaxy center has no
observed counterpart. In this respect, model~7 (Fig.~\ref{fig:N4388_final_test3_deltat1_plots_hi}) better reproduces the H{\sc i} observations.
The main drawback of the model (and all other models) is  the lack of emission at vertical distances $60 \leq z \leq 90$~kpc, $30$ to $50$~kpc
east of the galaxy center.
 
For the model pV diagram we assume that only the parts with the highest particle density can be observed.
With $\sim 500$~km\,s$^{-1}$, the velocity range of the model gas plume is well comparable to that of the observed plume (lower left panel of Fig.~\ref{fig:hicomp}). 
The width of the model tail in the pV diagram is about $100$~km\,s$^{-1}$, well comparable to the observed velocity width.
However, the smaller velocity width of the farthest parts of the tail is not reproduced by the model.

We conclude that galactic structure, i.e. the location of spiral arms during the ram pressure stripping event,
determines the distribution of the gas surfaces density within the extended gas plume to a certain extent (gas surface density near the galactic disk,
morphology of the vertical part). However, it cannot solve the discrepancy between the model and observations in the northwestern region of the gas tail
which is closest to M~86 in projection on the sky (Fig.~\ref{fig:hicomp}).
\begin{figure*}
  \resizebox{\hsize}{!}{\includegraphics{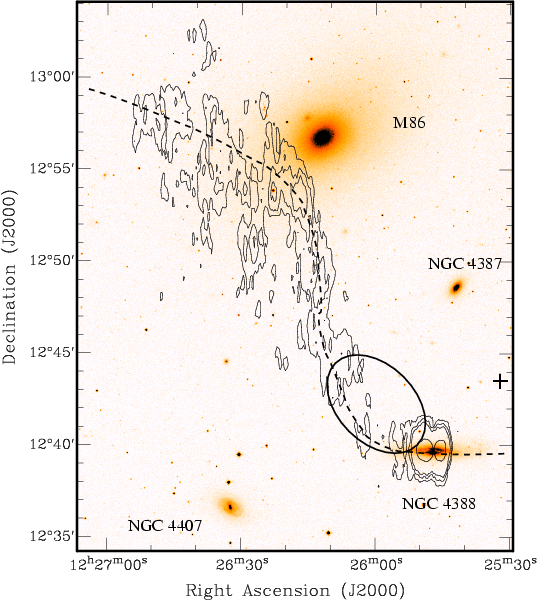}\includegraphics{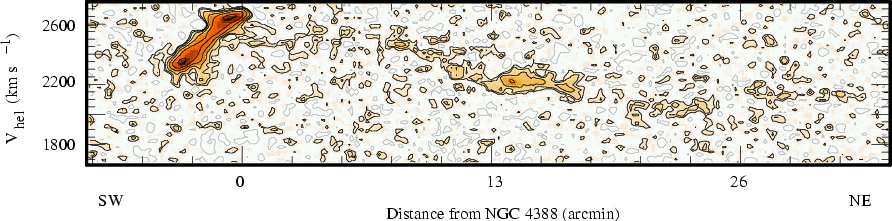}}
        \resizebox{\hsize}{!}{\includegraphics{N4388_final_test3_paper30n_6_HI.png}}
        \caption{Upper left panel: H{\sc i} on optical image. The contour levels are $(1,5,10,50) \times 10^{19}$~cm$^{-2}$. 
          Upper right panel: H{\sc i} pV diagram from Oosterloo \& van Gorkom (2005). 
          Lower left panel: ``best-fit'' model H{\sc i} large-scale distribution with $i=90^{\circ}$.  Contour levels are
          $(3,6,9,15,30,150) \times 10^{19}~{\rm cm}^{-2}$.
          Lower right panel: model H{\sc i} pV diagram.
        } \label{fig:hicomp}
\end{figure*}

\subsubsection{The H{\sc i} disk}

\begin{figure*}
        \resizebox{\hsize}{!}{\includegraphics{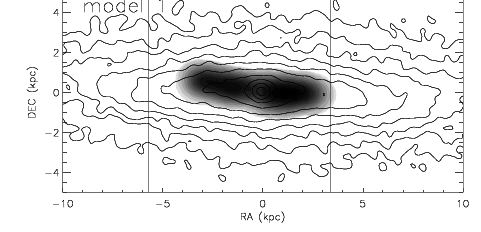}\includegraphics{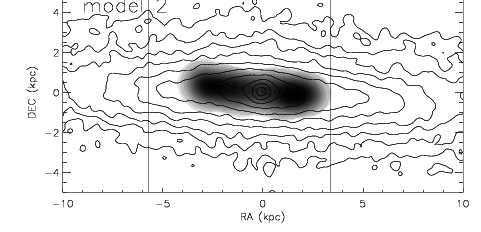}}
        \resizebox{\hsize}{!}{\includegraphics{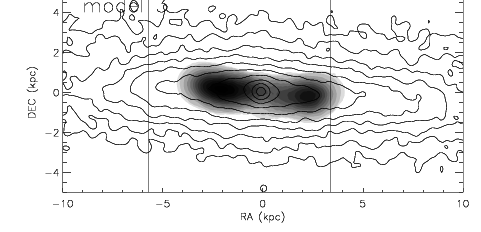}\includegraphics{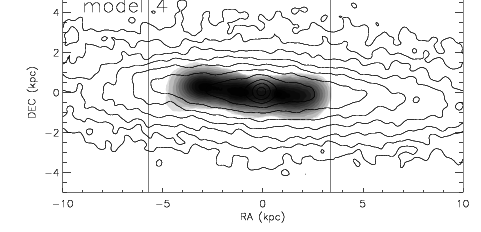}}
        \resizebox{\hsize}{!}{\includegraphics{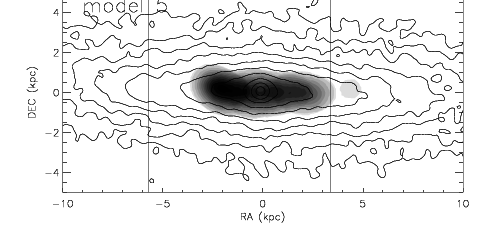}\includegraphics{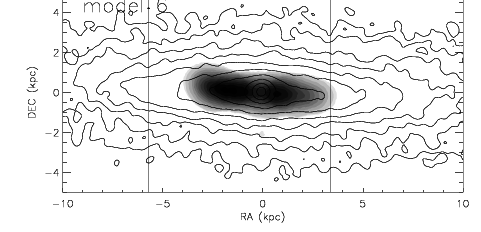}}
        \resizebox{\hsize}{!}{\includegraphics{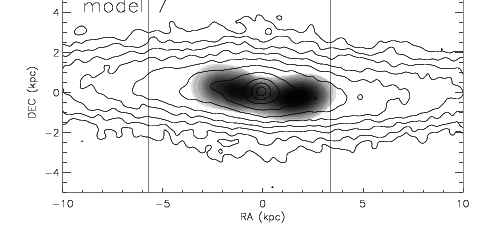}\includegraphics{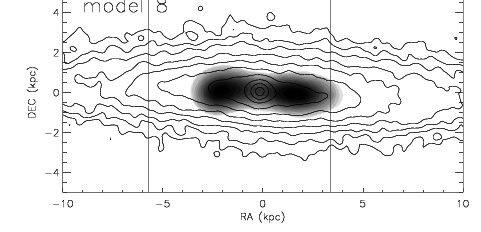}}
        \resizebox{\hsize}{!}{\includegraphics{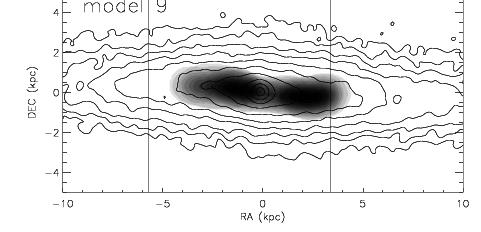}\includegraphics{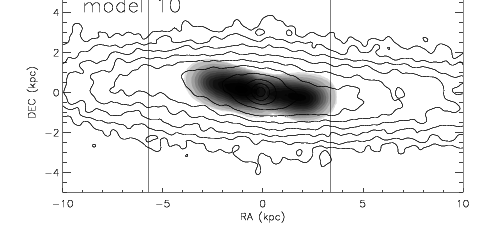}}
        \caption{Model H{\sc i} distributions of the galactic disk (greyscale) with $i=90^{\circ}$. 
          Greyscale levels are (4, 7, 14, 22, 30, 36, 43, 50, 58, 65, 72, 80)$\times 10^{20}$~cm$^{-2}$. The stellar content is shown with contours.
          To guide the eye, the vertical lines delimitate the extent of the observed H{\sc i} distribution.
        } \label{fig:N4388_final_test3_deltat1_plots_hi}
\end{figure*}

The observed and model H{\sc i} distributions within the galactic disk are presented in Fig.~\ref{fig:N4388_final_test3_deltat1_plots_hi} for
an inclination angle of $i=90^{\circ}$. Since the projection with $i=75^{\circ}$ does not significantly differ from the projection with $i=90^{\circ}$,
we do not show it in the Appendix.
The observed western extent of the H{\sc i} disk is reproduced by all models. However, a pronounced east-west asymmetry, as it is observed, is only present in
models~4 and 9. Still, the eastern half of the observed H{\sc i} disk is $20$\,\% and $25$\,\% more extended than those of model~4 and 9, respectively.

\begin{figure*}
        \resizebox{\hsize}{!}{\includegraphics{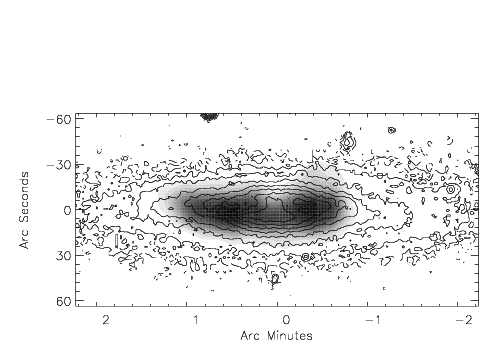}\includegraphics{N4388_final_test3_paper30n_6_plots_81_90_HI.png}}
        \caption{Left panel: VIVA H{\sc i} distribution (greyscale; Chung et al. 2009).
          Greyscale levels are (2, 4, 6,..., 40)$\times 10^{20}$~cm$^{-2}$.
          Right panel: model H{\sc i} distribution with $i=90^{\circ}$. Greyscale levels are (4, 7, 14, 22, 30, 36, 43, 50, 58, 65, 72, 80)$\times 10^{20}$~cm$^{-2}$.
        The stellar content is shown with contours.
        To guide the eye, the vertical lines delimit the extent of the observed H{\sc i} distribution.} \label{fig:hicomp1}
\end{figure*}

The observed and ``best-fit'' model H{\sc i} distributions within the galactic disk are presented in Fig.~\ref{fig:hicomp1}. 
Since we restricted ourselves to model~9 and 10 for the ``best-fit'' model, we present model~9 as the ``best-fit'' model.
The observed H{\sc i} distribution has a central 
depression and is truncated at a radius of $5.7$~kpc to the east and $3.4$~kpc to the west with respect to the optical center of the galaxy.
The model shows the same, but less pronounced, east-west asymmetry. However the eastern side of the model H{\sc i} disk is $\sim 1.5$~kpc less extended
than its observed counterpart. None of the models reproduces the observed protruding H{\sc i} in the northwestern part of the galactic disk.

\begin{figure}
        \resizebox{\hsize}{!}{\includegraphics{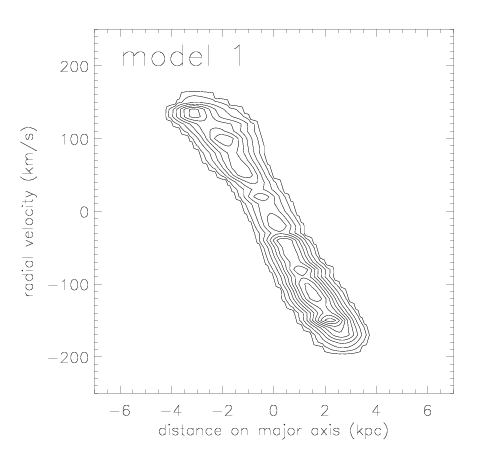}\includegraphics{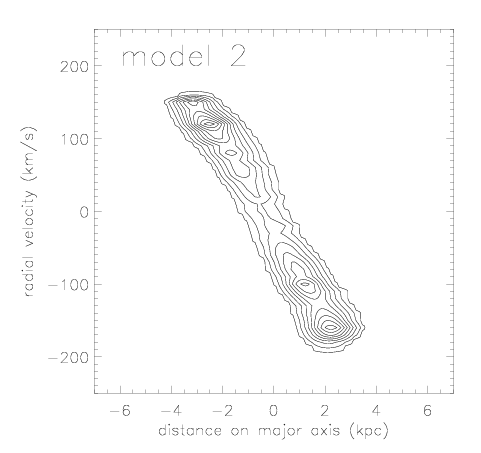}}
        \resizebox{\hsize}{!}{\includegraphics{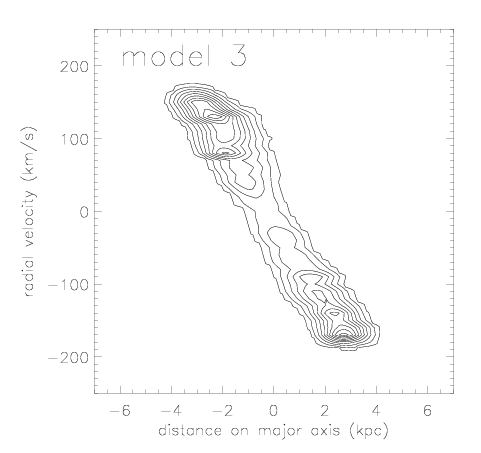}\includegraphics{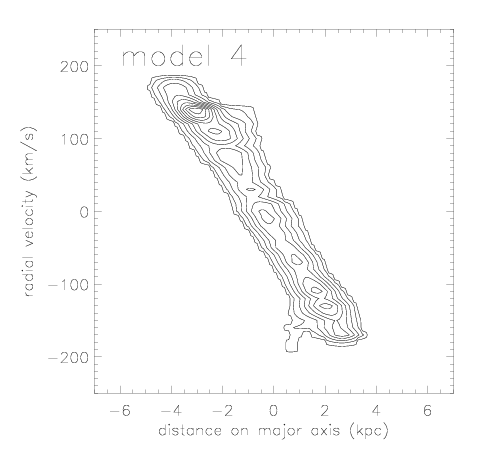}}
        \resizebox{\hsize}{!}{\includegraphics{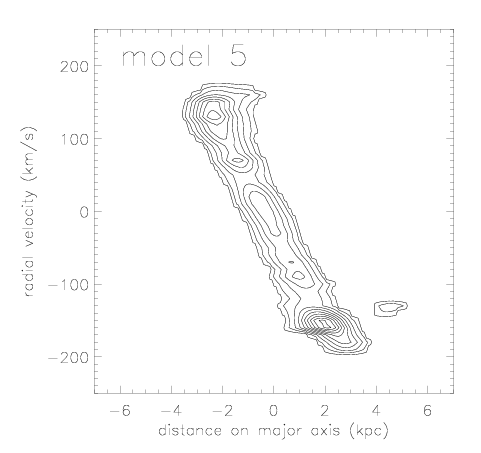}\includegraphics{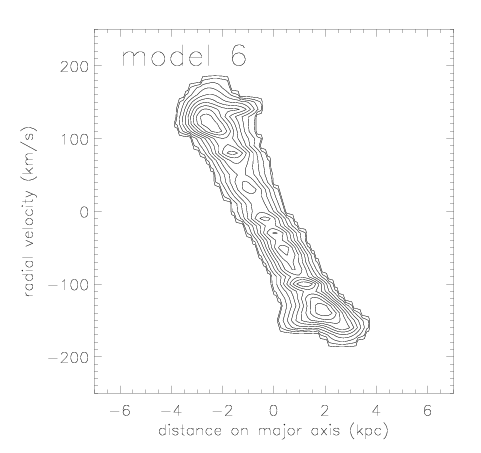}}
        \resizebox{\hsize}{!}{\includegraphics{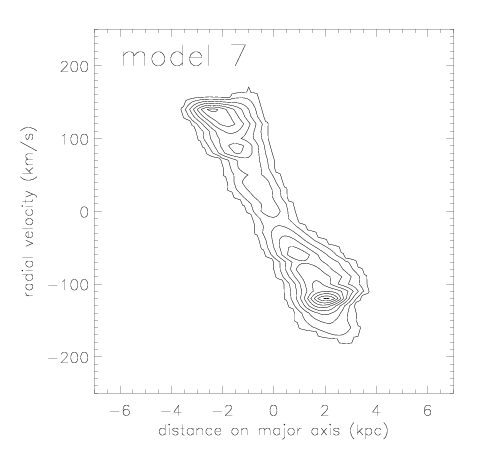}\includegraphics{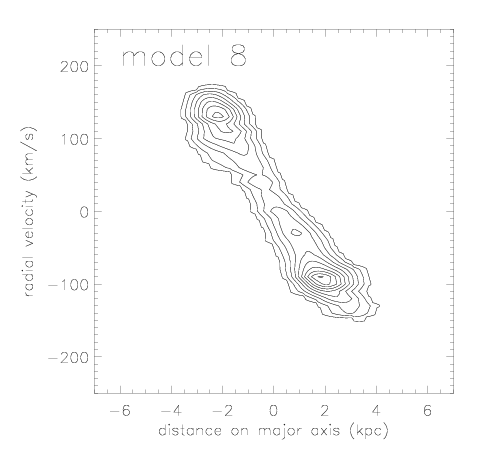}}
        \resizebox{\hsize}{!}{\includegraphics{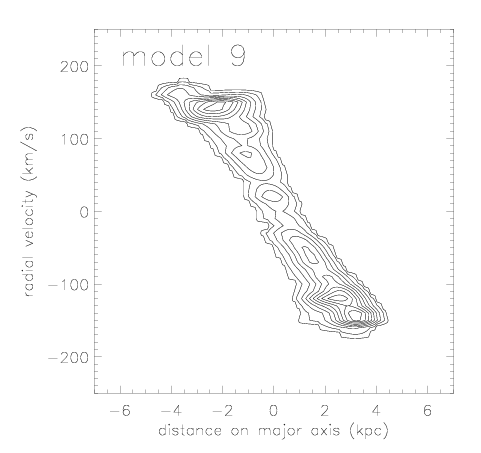}\includegraphics{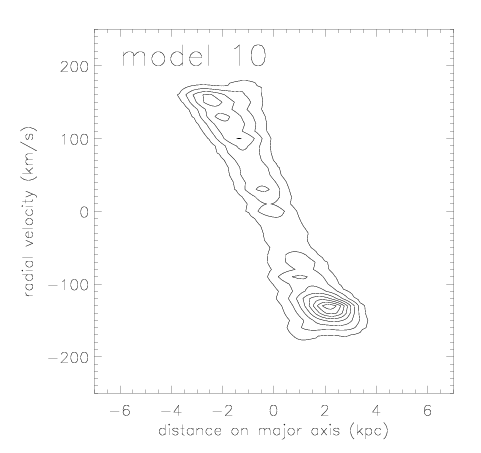}}
        \caption{Model H{\sc i} pV diagrams along the major axis with $i=90^{\circ}$.
        } \label{fig:N4388_final_test3_deltat1_plots_pv}
\end{figure}

The position-velocity diagrams along the galaxy's major axis are presented in Fig.~\ref{fig:N4388_final_test3_deltat1_plots_pv} for an inclination
angle of $i=90^{\circ}$. Since the projection with $i=75^{\circ}$ does not significantly differ from the projection with $i=90^{\circ}$,
we do not show it in the Appendix.
Only models~1, 7, 9, and 10 show a small plateau of constant rotation velocity at the eastern side of the galactic disk.
All simulation snapshots have in common that the western side of the pV diagram show a linear behaviour.

\begin{figure*}
        \resizebox{\hsize}{!}{\includegraphics{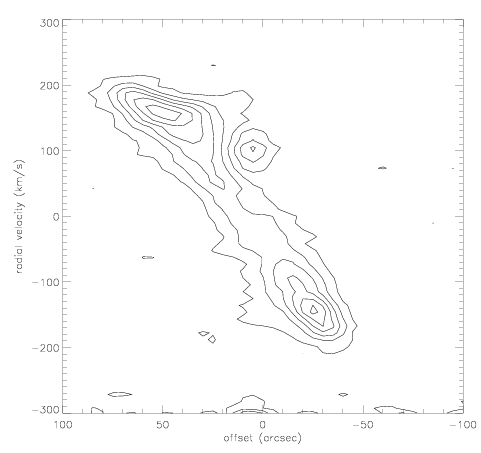}\includegraphics{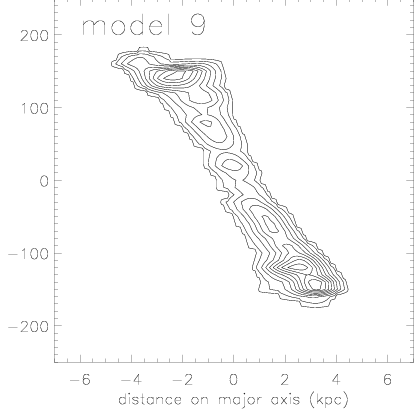}}
\caption{Left panel: VIVA H{\sc i} pV diagram along the major axis (Chung et al. 2009). 
  Right panel: model H{\sc i} pV diagram along the major axis with $i=90^{\circ}$.} \label{fig:hicomp2}
\end{figure*}

The observed position-velocity diagram shows a different kinematical behaviour of the two sides of the galactic disk (left panel of Fig.~\ref{fig:hicomp2}):
the western side has an almost constant velocity gradient until the outer edge of the disk, whereas the velocity gradient of the eastern side
decreases significantly at a distance of $> 40'' = 3.4$~kpc. This behaviour is qualitatively best reproduced by model~9 (right panel of Fig.~\ref{fig:hicomp2}).

We conclude that galactic structure has a small but significant influence on the east-west asymmetry of the truncated H{\sc i} disk.

\subsection{H$\alpha$}

\begin{figure*}
        \resizebox{\hsize}{!}{\includegraphics{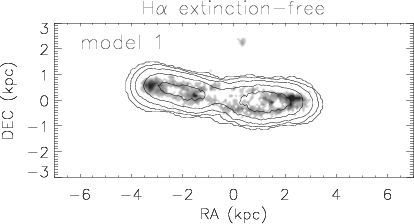}\includegraphics{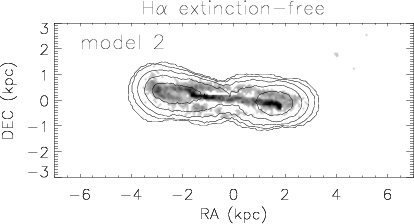}}
        \resizebox{\hsize}{!}{\includegraphics{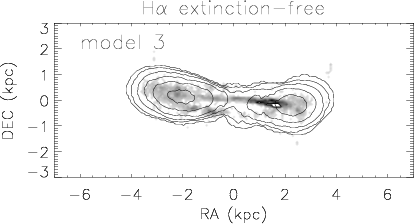}\includegraphics{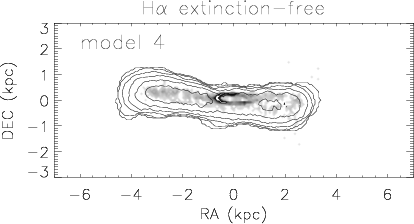}}
        \resizebox{\hsize}{!}{\includegraphics{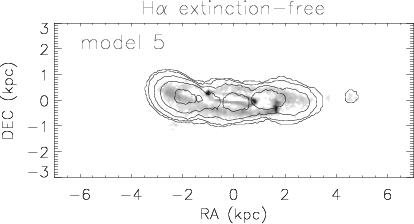}\includegraphics{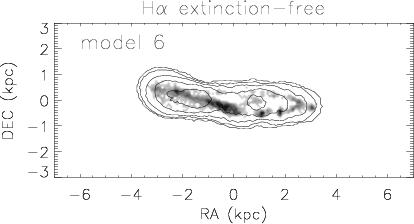}}
        \resizebox{\hsize}{!}{\includegraphics{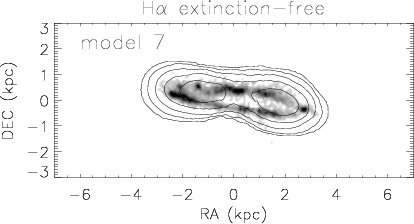}\includegraphics{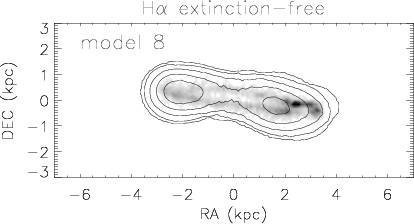}}
        \resizebox{\hsize}{!}{\includegraphics{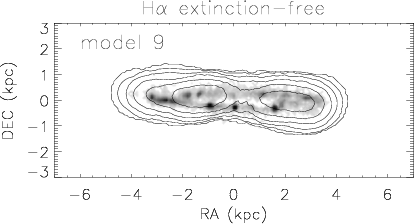}\includegraphics{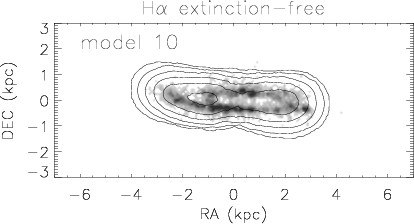}}
        \caption{Model H{\sc i} contours on model H$\alpha$ emission distributions without extinction with $i=75^{\circ}$.
        } \label{fig:N4388_final_test3_deltat1_plots_ha}
\end{figure*} 
The model H$\alpha$ maps of all models are presented in Fig.~\ref{fig:N4388_final_test3_deltat1_plots_ha} for an inclination angle of $i=75^{\circ}$.
The corresponding model images for $i=90^{\circ}$ are shown in  Fig.~\ref{fig:N4388_final_test3_deltat1_plots_ha1}.
By comparing the outer spiral arms (especially in models 6, 9, and 10) between the two figures, it becomes clear that the inclination angle changes the apparent sense of rotation:
whereas it is counter-clockwise for $i=75^{\circ}$, it is clockwise for $i=90^{\circ}$. Assuming trailing spirals, observations indicate
an apparent  counter-clockwise rotation of the galactic disk. All model H$\alpha$ distribution show a central ring structure.
In most cases the ring has an asymmetric surface brightness distribution.
In addition, models~6, 9, and 10 show an arm structure in the eastern part of the galactic disk.

\begin{figure*}
        \resizebox{\hsize}{!}{\includegraphics{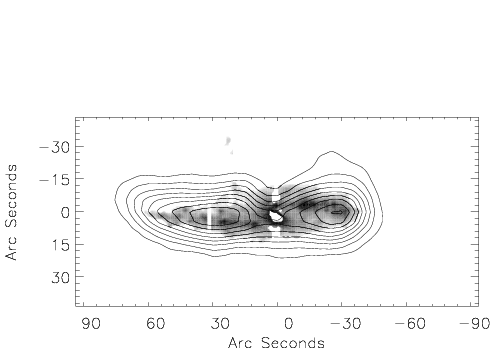}\includegraphics{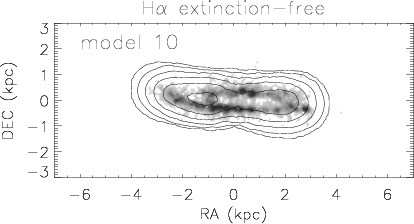}}
\caption{Left panel: VIVA H{\sc i} contours on H$\alpha$ emission distribution (Yoshida et al. 2002). 
         Right panel: model H{\sc i} contours on model H$\alpha$ emission distribution without extinction with $i=75^{\circ}$.} \label{fig:hacomp1}
\end{figure*}

The observed and ``best-fit'' model (model~10) H$\alpha$ maps are presented in Fig.~\ref{fig:hacomp1}. 
The Spitzer $8$~$\mu$m emission distribution is comparable to the H$\alpha$ emission.
These maps show the same east-west asymmetry as the H{\sc i} distribution: the eastern side is $\sim 1.5$ times  more extended than the western side.
Two prominent spiral arms are visible in the northwestern and southeastern quadrants of the galactic disk.
On the other hand, the $6$~cm radio continuum map (Fig.~2 of Damas-Segovia et al. 2016) shows more a ring structure in the western part of the galactic disk.
Thus, a high extinction is not excluded in the southwestern quadrant of the galactic disk.
The only model which resembles the H$\alpha$/$8$~$\mu$m emission distribution of the eastern half of the galactic disk is model~10. 
Only in the presence of high extinction in the southwestern quadrant of the galactic disk, the H$\alpha$/$8$~$\mu$m emission distribution of the 
western half of the galactic disk is reproduced by the model.

\subsection{FUV}

\begin{figure*}
        \resizebox{\hsize}{!}{\includegraphics{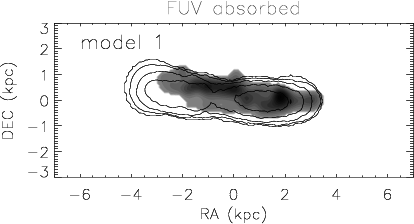}\includegraphics{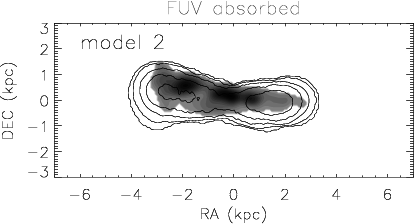}}
        \resizebox{\hsize}{!}{\includegraphics{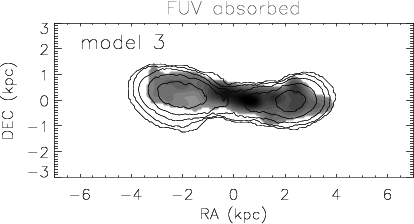}\includegraphics{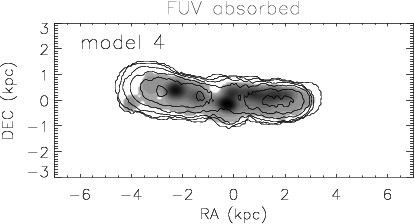}}
        \resizebox{\hsize}{!}{\includegraphics{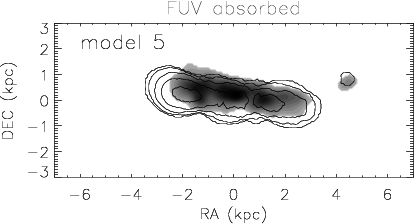}\includegraphics{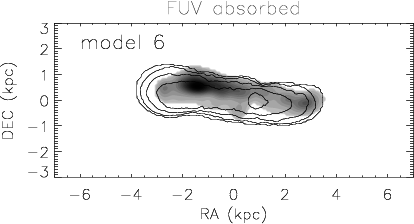}}
        \resizebox{\hsize}{!}{\includegraphics{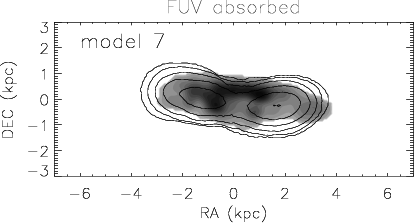}\includegraphics{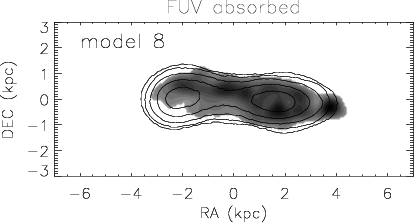}}
        \resizebox{\hsize}{!}{\includegraphics{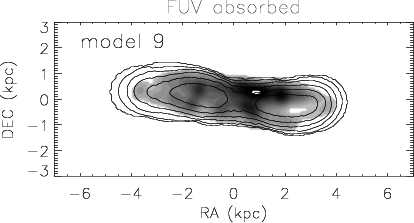}\includegraphics{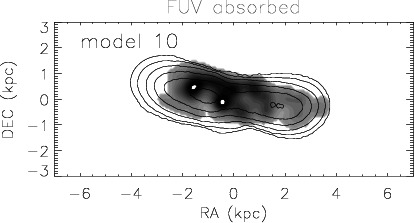}}
        \caption{Model H{\sc i} contours on model FUV emission distributions with $i=90^{\circ}$.
        } \label{fig:N4388_final_test3_deltat1_plots_fuv}
\end{figure*}

\begin{figure*}
        \resizebox{\hsize}{!}{\includegraphics{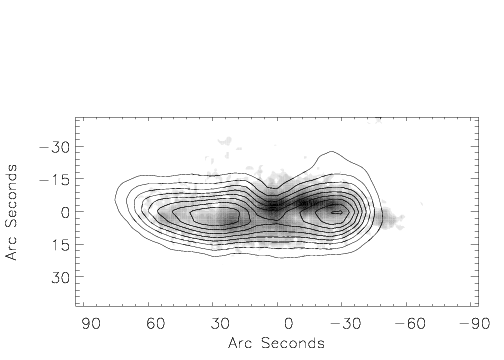}\includegraphics{N4388_final_test3_paper30n_6_plots_81_90_fuvabs.png}}
\caption{Left panel: VIVA H{\sc i} contours on GALEX FUV emission distribution. Right: model H{\sc i} contours on model FUV emission distribution with $i=90^{\circ}$.} \label{fig:fuvcomp}
\end{figure*}

The FUV emission distribution traces the less recent star formation ($\sim 100$~yr) than the H$\alpha$ emission distribution ($\sim 10$~Myr).
In addition, the FUV emission is prone to dust extinction. We determined the model 3D FUV emission by using the clocks associated to the newly
created stellar particles (see Sect.~\ref{sec:sfr}). In a second step, we calculated the dust extinction based on the 3D gas distribution
by assuming a constant gas-to-dust ratio of $100$. The resulting observed and model FUV emission distributions are presented in Fig.~\ref{fig:N4388_final_test3_deltat1_plots_fuv}
for an inclination of $i=90^{\circ}$, those for $i=75^{\circ}$ are shown in Fig.~\ref{fig:N4388_final_test3_deltat1_plots_fuv1}.
For comparison, the GALEX FUV emission of NGC~4388 (left panel of Fig.~\ref{fig:fuvcomp}) is very asymmetric: 
only the northwestern spiral arm is clearly visible. This means that the southeastern spiral arm is strongly obscured by the disturbed dusty interstellar medium of NGC~4388.

The model FUV emission distribution of almost all models, except model~3 with $i=90^{\circ}$, are asymmetric. Only the asymmetry of model~9 with $i=90^{\circ}$
is comparable to the observed asymmetry, i.e. the emission in the northwestern quadrant of the galactic disk is prominent. 
We therefore assume that model~9 with $i=90^{\circ}$ reproduces ``best'' the GALEX FUV observations.

\subsection{Polarized radio continuum}

To calculate the large-scale regular magnetic field for our simulations,
we follow the same procedure as Soida et al. (2006). Time-dependent gas-velocity
fields provided by the 3D dynamical simulations were used as the input for 
solving the dynamo (induction) equation using a second order Godunov
scheme with second order upstream partial derivatives together with a second
order Runge-Kutta scheme for the time evolution. To avoid non-vanishing 
$\nabla\cdot\vec B$, we evolved the dynamo equation expressed by the magnetic 
potential $\vec A$, where  $\vec B=\nabla\times\vec A$, ($\vec B$ is the
magnetic induction):
\begin{equation}
{\partial \over \partial t} \vec B = \nabla\times(\vec v \times \vec B )
 \nabla\times (\alpha~\vec B) - \nabla\times(\eta~\nabla\times \vec B)
\label{eq:inductioneq}
\end{equation}
where $\vec v$ is the large-scale velocity of the gas, and $\eta$ the
coefficient of a turbulent diffusion. We assume the magnetic field to be
partially coupled to the gas via the turbulent diffusion process
(Elstner et al. 2000) assuming the magnetic diffusion coefficient to be
$\eta = 3\times 10^{25} {\rm cm}^2{\rm  s}^{-1}$. We do not implement any explicit dynamo
process ($\alpha = 0$).
Our MHD calculations were carried out on 3D grid of $256^3$ points, and 
with 300\,pc spatial resolution. For the initial conditions we adopt
the magnetic vector potential appropriate to purely toroidal magnetic
field with an exponential distribution with a scalelength of 1 kpc in the
vertical direction.
The MHD model does not contain a galactic wind.

The resulting polarized emission is calculated by assuming a density of relativistic electrons that is
proportional to the model gas density $\rho$. This rather crude approximation is motivated by the fact that
in quiescent galaxies, the density of relativistic electrons is approximately proportional to the star formation density
which depends on $\rho^{1{\rm -}1.7}$.

\begin{figure*}
        \resizebox{16cm}{!}{\includegraphics{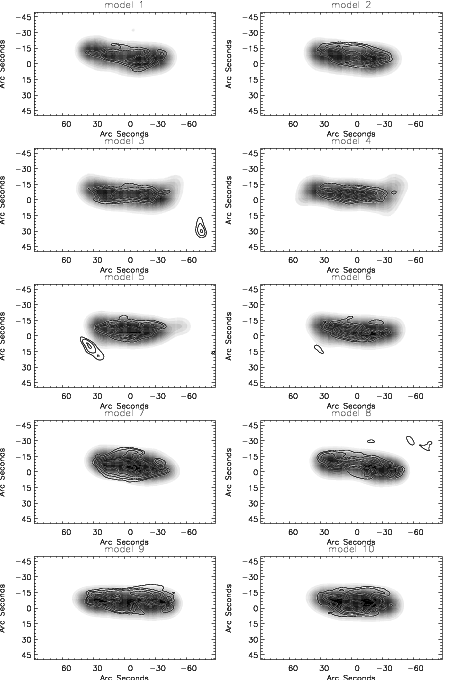}}
        \caption{Model polarized emission distributions on model H{\sc i} distribution with $i=75^{\circ}$.
        } \label{fig:pivergleich1}
\end{figure*}

The model and observed 6~cm polarized radio continuum emission distributions are presented in Fig.~\ref{fig:pivergleich1} for an inclination angle of $i=75^{\circ}$.
The corresponding maps for $i=90^{\circ}$ are shown in Fig.~\ref{fig:pivergleich1a}.
All model 6~cm polarized radio continuum emission distributions, except that of model~6 with $i=75^{\circ}$, are asymmetric.
A significant east-west asymmetry at $i=75^{\circ}$ is observed in model~1, 2, 6, and 10, at $i=90^{\circ}$ in model~1 and 3.
A significant north-south asymmetry at $i=75^{\circ}$ is observed in model~10, at $i=90^{\circ}$ in model~2, 7 and 10.

\begin{figure*}
        \resizebox{\hsize}{!}{\includegraphics{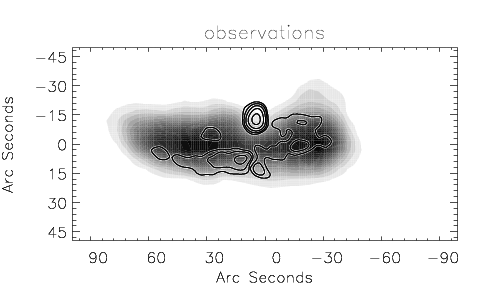}\includegraphics{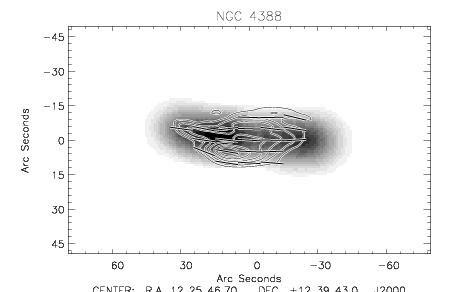}}
        \caption{Left panel: 6~cm polarized emission distribution (Damas-Segovia et al. 2016) on VIVA H{\sc i} distribution. 
          Right panel: model polarized emission distribution on model H{\sc i} distribution with $i=75^{\circ}$.} \label{fig:pivergleich1b}
\end{figure*}

The observed polarized emission at $6$~cm is asymmetric (left panel of Fig.~\ref{fig:pivergleich1b}). Whereas it is shifted
to the south with respect to the galactic plane in the eastern half, it is shifted to the north in the western half of the galactic disk.
The observed polarized emission along the minor axis of the galactic disk is caused by the nuclear outflow (Damas-Segovia et al. 2016; see also 
Fig.~5 of Yoshida et al. 2002).

The only model with a significant north-south asymmetry in the eastern half of the galactic disk is model~10 at an inclination of $i=75^{\circ}$
(right panel of Fig.~\ref{fig:pivergleich1b}). It also shows some emission in the northwestern quadrant of the galactic disk, as it is observed.
However, the model polarized radio continuum emission has a significant disk component which is not observed.

\subsection{Optical spectrum of the gas-free part of the galactic disk}

We used the VLT FORS2 spectrum from Pappalardo et al. (2010) obtained with the GRIS-600B covering a wavelength range of $3350$-$6330$~{\AA} with a resolution
of $1.48$~{\AA} binned pixel. The average signal-to-noise ratio per pixel is $\sim 26$.

\begin{figure}
        \resizebox{\hsize}{!}{\includegraphics{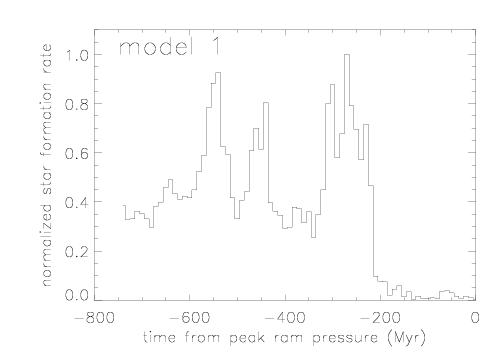}\includegraphics{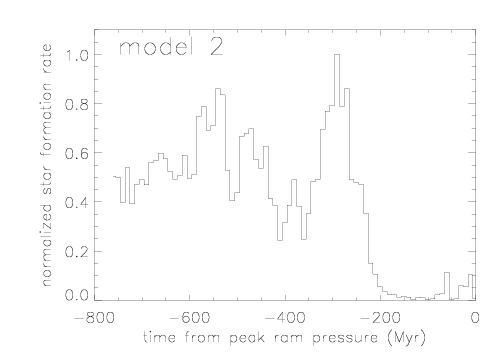}}
        \resizebox{\hsize}{!}{\includegraphics{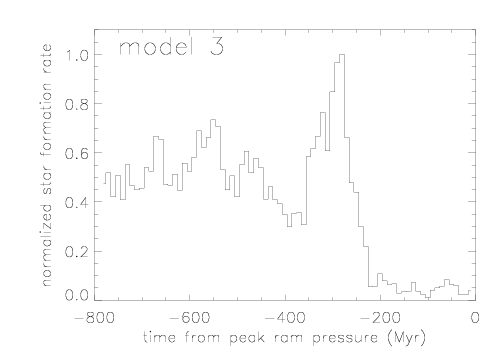}\includegraphics{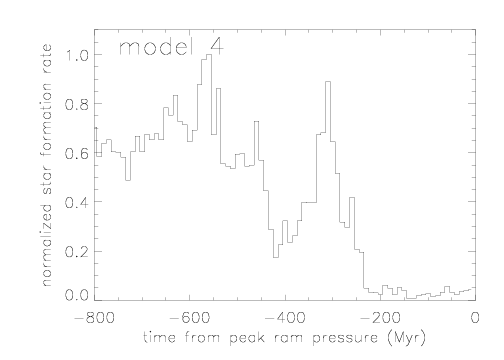}}
         \resizebox{\hsize}{!}{\includegraphics{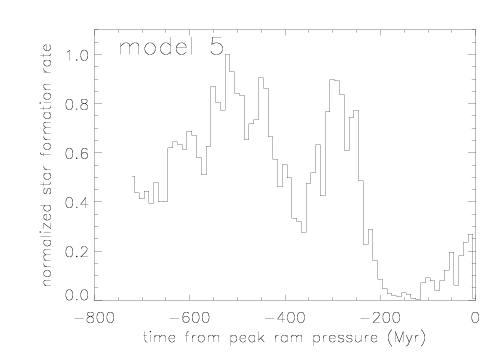}\includegraphics{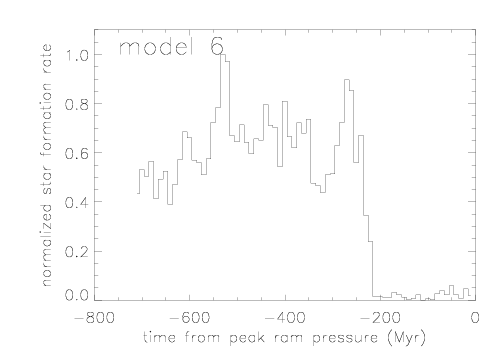}}
        \resizebox{\hsize}{!}{\includegraphics{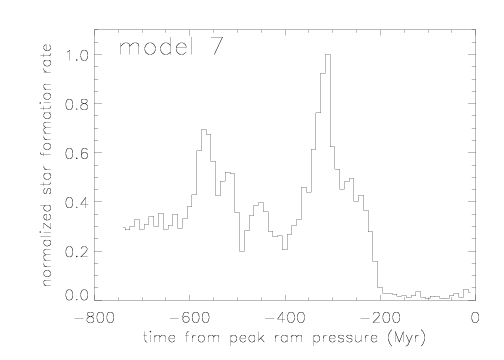}\includegraphics{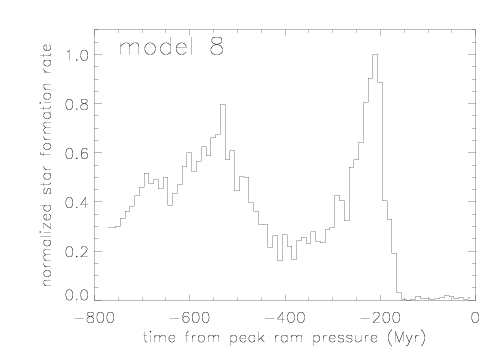}}
        \resizebox{\hsize}{!}{\includegraphics{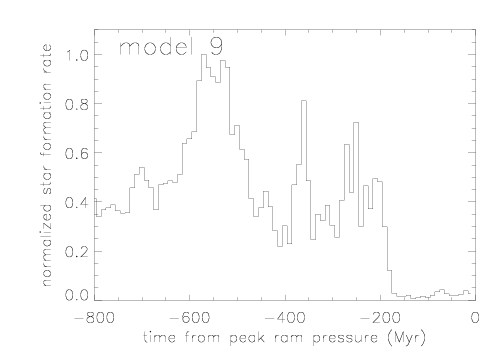}\includegraphics{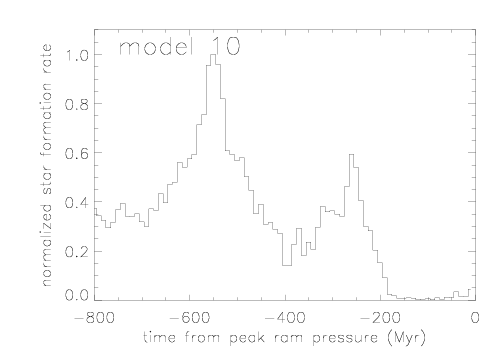}}
        \caption{Normalized model star formation histories. The time $t= 0$~Myr is the time of interest.
        } \label{fig:N4388_final_test3_deltat1_plots_sfh} 
\end{figure}

As in Pappalardo et al. (2010) we derived an optical spectrum from our simulations within a slit which is located at a distance of $4.5$~kpc west of the galaxy center.
The slit size is $500$~pc. To do so, we extracted all star particles from this regions to obtain the star formation histories within
this region for our ten models (Fig.~\ref{fig:N4388_final_test3_deltat1_plots_sfh}). The variations of the star formation rate for $t < -200$~Myr
is due to galactic spiral arms. From the beginning of the simulations the model galaxy is affected by ram pressure which leads to
an asymmetric distribution of the outer gas disk after $\sim 150$--$200$~Myr. Even small differences in the amount of ram pressure between the models at the beginning
of the simulations with different time delays lead to a different star formation history within the slit at the times of interest.
Models~3, 7, and 8 show an increase of the star formation rate just before gas stripping that is more
important than the increase caused by spiral arms. The star formation of all models with a stripping angle of $45^{\circ}$ steeply decrease around 
$t \sim -250$~Myr due to gas stripping by ram pressure. This timescale is significantly higher than the star formation quenching time of $t=-190 \pm 30$~Myr obtained 
by Pappalardo et al. (2010). For a stripping angle of $30^{\circ}$, star formation strongly decreases at $t \sim -200$~Myr.

\begin{figure*}
        \resizebox{\hsize}{!}{\includegraphics{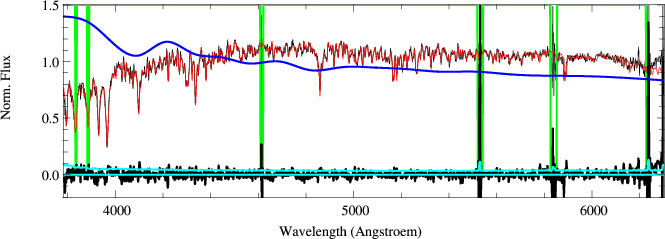}}
        \caption{Optical spectrum for model~9 (red) with Z=0.014, $fct=1.2$, $\tau=70$~Gyr together with the observed spectrum. The residuals are shown as the lower black line.
          Blanked regions are shown in green. The bottom of the panel shows the residuals (black) and the observational errors (cyan). 
          The subtracted continuum is shown as a blue line.
        } \label{fig:bvmod4_tau2000_z0.014_fct1.0_newfct_spectra}
\end{figure*}

\begin{figure}
        \resizebox{\hsize}{!}{\includegraphics{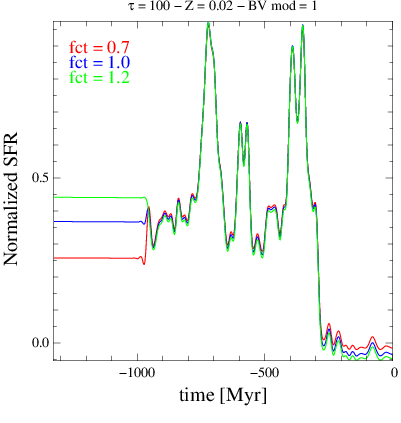}}
        \caption{Star formation histories of model~1 for which model spectra were obtained. The timescale for the exponential star formation rate is
            $\tau=100$~Gyr (Eq.~\ref{eq:tau}).
        } \label{fig:sfh_bvmod1_tau1000_fct_testv2}
\end{figure}

We used the star formation histories of Fig.~\ref{fig:N4388_final_test3_deltat1_plots_sfh} to obtain optical spectra for the different models.
Only one of these spectra is shown in Fig.~\ref{fig:bvmod4_tau2000_z0.014_fct1.0_newfct_spectra}, because of the strong resemblance between the spectra.
Since the model star formation histories end at $t=-800$~Myr we had to assume a star formation history prior to this timestep.
We decided to use an exponentially declining star formation rate with three different normalizations (Fig.~\ref{fig:sfh_bvmod1_tau1000_fct_testv2})
\begin{equation}
\label{eq:tau}
SFR = fct \times SFR_0 \times \exp (-t/\tau)\ ,
\end{equation}
where $fct=0.7,1,1.2$ and $\tau=20,25,30,40,70,100,200,400$~Gyr.
The range of metallicities was set to $Z =0.01$ to $0.028$ in steps of $0.02$. The revised solar metallicity is $Z =0.014$ (Asplund et al. 2009).
For each comparison of the model spectrum with the observed spectrum we calculated an associated $\chi^2$.
In addition, we only allowed for models whose $\chi^2$ at the time of interest is minimum, i.e. the $\chi^2$ increases for earlier or later
times of interest. Without doing so, a discrimination between the models based on the $\chi^2$ distribution was not possible. This means that at this
point of a refined modelling the optical spectrum gives only weak constraints.

\begin{figure}
        \resizebox{\hsize}{!}{\includegraphics{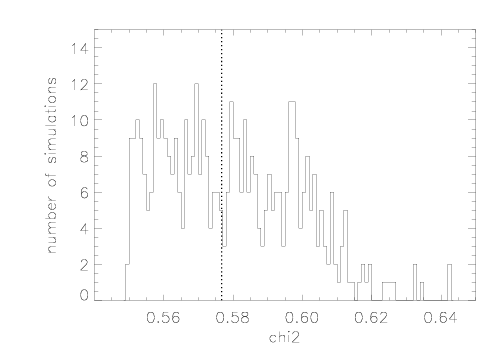}}
        \resizebox{\hsize}{!}{\includegraphics{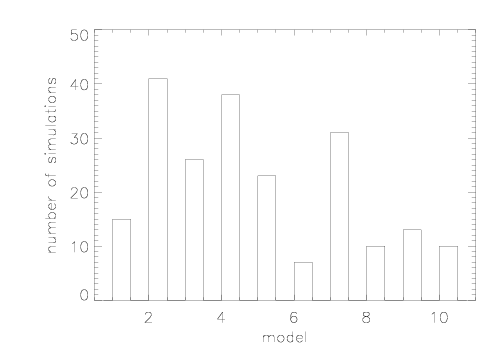}}
        \caption{Upper panel: distribution of the model $\chi^2$ for $Z \geq 0.01$. The dotted line represents the
          limit ($1.05 \times {\rm min}(\chi^2)$) for the accepted models shown in the lower panel. 
          Lower panel: the distribution of the different models with the lowest $\chi^2$.
        } \label{fig:ki2}
\end{figure}

The distributions of the $\chi^2$ for the different model spectra are shown in Fig.~\ref{fig:ki2} for the solar and slightly subsolar metallicity models.
In general, the most important influence on the $\chi^2$ values comes from metallicity: low metallicity models yield smaller $\chi^2$ than high metallicity models.

The percentage for the acceptance of a model $\chi^2$ is set to the inverse of the square root of the number of degrees of freedom.
The number of degrees of freedom $N$ is set to the number of independent points of the spectrum. With a spectral resolution of 
$\Delta \lambda / \lambda=780$, a wavelength coverage of $2400 \AA$, and a central wavelength of $5000 \AA$, we find $N \sim 374$.
We thus set the limit for the acceptance of a model to $\chi^2_{\rm lim}=1.05 \times {\rm min}(\chi^2)$.

The smallest $\chi^2$ value is $\chi^2=0.549$ at a metallicity $Z=0.01$. Models with a normalization $fct=0.7$ do not appear in the list of acceptable models.
Moreover, models with $20 \leq \tau \leq 40$~Gyr are preferred. 
For their parametric determination of the star formation quenching time Pappalardo et al. (2010) found a minimum value of 
$\chi^2=0.597$ for $Z=0.018$. The corresponding $\chi^2$ of model~9 with $Z=0.018$ is $\chi^2=0.593$, i.e. the
star formation history of model~9 leads to a slightly better fit of the optical spectrum than a truncated constant star formation rate.
The distributions of the different models with lowest $\chi^2$ are shown in the lower panel of Fig.~\ref{fig:ki2}.
For both metallicities, $Z=0.014$ and $Z=0.01$, models~2, 3, 4, 5, and 7 are preferred, whereas models~1, 6, 8, 9, and 10 are underrepresented.

We define the model star formation quenching age as the time when the local star formation rate dropped below $1/3$ of
its mean value for $t < -230$~Myr.
The obtained star formation quenching age of $190$~Myr and $200$~Myr for models~9 and 10 are consistent with the quenching age
derived by Pappalardo et al. (2010) within their uncertainty of $30$~Myr. Our quenching age
is also broadly consistent with the quenching age determined by Crowl et al. (2008) from spectroscopic line indices.
Given the differences in the model star formation histories before gas stripping, we believe that it is not excluded that 
Pappalardo et al. (2010), who assumed smooth star formation histories, have somewhat underestimated the uncertainty of the quenching age.

\section{Discussion}

Table~\ref{tab:results} summarizes the results obtained in Sect.~\ref{sec:comparison}.
Models~1 to 8 do not reproduce the majority of the available observations.
Thus, we conclude that models~9 and 10 are the preferred models and derive a ram pressure wind angle of $30^{\circ}$ from our simulations.
Even the preferred models have difficulties to reproduce the detailed large-scale H{\sc i}
emission distribution and the polarized radio continuum emission distribution.  
The preferred inclination angle is $i=75^{\circ}$, except for the FUV emission distribution which is only reproduced with $i=90^{\circ}$.
Since this inclination leads to an apparent direction of rotation which is opposite to the observed direction, 
$i=75^{\circ}$ is the best choice for the inclination. For comparison, Veilleux et al. (1999) gave an inclination angle of $i=78^{\circ}$.

Models~9 and 10 reproduce the extent and the radial velocity range of the H{\sc i} tail. Its width is only reproduced in the southern half of the tail.
These models also reproduce the east-west asymmetry of the galactic H{\sc i} disk, its pV diagram, the FUV emission distribution, and the optical spectrum.
Moreover, they qualitatively reproduce the main features of the H$\alpha$ and polarized radio continuum emission distributions.
However, these ``best-fit'' models do not reproduce the morphology of the northern half of the H{\sc i} tail, the protruding H{\sc i} feature in the
northwestern part of the galactic disk, the smaller velocity width of the far tail, and the detailed H$\alpha$ and radio continuum emission distributions.  
\begin{table*}
      \caption{Comparison between the models and observations}
         \label{tab:results}
         \begin{tabular}{lcccccccccc}
           \hline
           model & 1 & 2 & 3 & 4 & 5 & 6 & 7 & 8 & 9 & 10 \\
           wind angle & $45^{\circ}$ & $45^{\circ}$ & $45^{\circ}$ & $45^{\circ}$ & $45^{\circ}$ & $45^{\circ}$ & $30^{\circ}$ & $30^{\circ}$ & $30^{\circ}$ & $30^{\circ}$ \\
           time of peak ram pressure $t_{\rm peak}$ (Myr) & 530 & 550 & 570 & 590 & 510 & 500 & 500 & 530 & 560 & 590 \\ 
           \hline
           large-scale HI & -- & -- &  -- & -- & -- & -- & $\sim$ & $\sim$ & $\sim$ & $\sim$ \\
           small-scale HI & -- & -- & -- & + & -- & -- & -- & -- & $\sim$ & -- \\
           HI pV & -- & -- & -- & $\sim$ & -- & -- & $\sim$ & -- & + & $\sim$ \\
           H$\alpha$/8~$\mu$m ($i=75^{\circ}$) & -- & -- & -- & -- & -- & -- & -- & -- & -- & $\sim$ \\
           H$\alpha$/8~$\mu$m ($i=90^{\circ}$) & -- & -- & -- & -- & -- & -- & -- & -- & -- & -- \\
           FUV ($i=75^{\circ}$) & -- & -- & -- & -- & -- & -- & -- & -- & -- & -- \\
           FUV ($i=90^{\circ}$) & -- & -- & -- & -- & -- & -- & -- & -- & + & -- \\
           polarized radio continuum ($i=75^{\circ}$) & -- & -- & -- & -- & -- & -- & -- & -- & -- & $\sim$ \\
           polarized radio continuum ($i=90^{\circ}$) & -- & -- & -- & -- & -- & -- & -- & -- & -- & -- \\
           optical spectrum & $\sim$ & + & + & + & + & $\sim$ & + & $\sim$ & $\sim$ & $\sim$  \\
           \hline
         \end{tabular} 
\end{table*}

The dependence on the time of peak ram pressure made clear that galactic structure, i.e. spiral arms, influence the outcome
of a ram pressure stripping event: the large-scale and small-scale H{\sc i} distribution and
the $\alpha$/8~$\mu$m, FUV, and polarized radio continuum emission distribution.  
This influence is similar to what was observed by Nehlig et al. (2016) for another Virgo galaxy, NGC~4501, which undergoes
active ram pressure stripping. 
The optical spectrum of the gas-free region of the galactic disk shows a rather weak dependence on galactic structure.
We thus conclude that for the detailed reproduction of multi-wavelength
observations of a spiral galaxy that undergoes or underwent a ram pressure stripping event,
galactic structure has to be taken into account.

The H{\sc i} disk truncation radius is given by the peak ram pressure during the stripping event ($p_{\rm max}=8.3 \times 10^{-11}$~dyn\,cm$^{-2}$).
The extent of the tail is determined by the width of the temporal ram pressure profile ($\Delta t = 100$~Myr).
This is entirely consistent with a radial orbit of NGC~4388 within the Virgo cluster (Vollmer et al. 2001). The present day
ram pressure is $p_{\rm today}=0.15 \times p_{\rm max} = 1.25 \times 10^{-11}$~dyn\,cm$^{-2}$. This value is consistent with the model
of Vollmer \& Huchtmeier (2003). With an intracluster medium density of $n_{\rm ICM}=3 \times 10^{-4}$~cm$^{-3}$ (Damas-Segovia et al. 2016)
and a molecular weight of $\mu=0.6$, this yields a galaxy velocity of $v_{\rm gal} \sim 2200$~km\,s$^{-1}$, which is about a factor of $1.5$
higher than its radial velocity with respect to the cluster mean. The normalized 3D vector of the galaxy velocity is
$\vec{v}_{\rm gal}^{\rm model}=(0.11,-0.69,0.72)$\footnote{Positive is west, north, and away from the observer.}. 
The model radial velocity is thus $v_{\rm rad}^{\rm model}=0.72 \times v_{\rm gal}^{\rm model}=1580$~km\,$s^{-1}$,
which is consistent with the observed radial velocity of $v_{\rm gal}=1420$~km\,$s^{-1}$.

At the time of interest, $240$~Myr after peak ram pressure, the disk gas is no longer actively stripped.
The inspection of the face-on evolution of the gas disk shows that already more than $100$~Myr before the time of interest
the truncated gas disks become oval and develop intermittent gas arms. Since there is no longer an overdensity
observed at the windward side, these structures are caused by the resettling of the gas that has been 
violently pushed to smaller galactic radii by strong ram pressure around its peak value.
In agreement with Damas-Segovia et al. (2016), we thus conclude that the observed asymmetries in the disk of NGC~4388 are not caused by
the present action of ram pressure, but by the resettling of gas that has been pushed out of the galactic disk during
the ram pressure stripping event.

With an H{\sc i} deficiency of $def_{\rm HI}=1.1 \pm 0.3$ and an H{\sc i} mass of $3.7 \times 10^{8}$~M$_{\odot}$ (Chung et al. 2009),
the total H{\sc i} mass prior to the ram pressure stripping event was of the order of $M_{\rm HI,\ prior} \sim 4 \times 10^9$~M$_{\odot}$ with an uncertainty of 
a factor of $2$.
Thus, a gas mass of $3.6 \times 10^9$~M$_{\odot}$ was stripped from the galaxy, of which only $3.4 \times 10^8$~M$_{\odot}$ or $10$\,\% were detected
by Oosterloo \& van Gorkom (2005). 
The two regions of highest column densities ($\sim 10^{20}$~cm$^{-2}$) in the model gas tail are located at projected coordinates $(-22~{\rm kpc},40~{\rm kpc})$
and $(-57~{\rm kpc},77~{\rm kpc})$. The observed regions of highest column densities ($\sim 10^{20}$~cm$^{-2}$) are located between $27$ and $36$~kpc
east and between $40$ and $70$~kpc north of the galaxy center. As predicted by Vollmer et al. (2001), even molecular gas clouds with masses of
a few $10^6$~M$_{\odot}$ exist in the plume region of high column densities at $z \sim 70$~kpc (Verdugo et al. 2015).

The initial radial profile of the gas surface density is expected to play a role for the morphology of the H{\sc i} tail
at the time of interest. Our initial model profile shows a rapid decrease from $10$ to $4$~M$_{\odot}$pc$^{2}$ between $7.5$~kpc and $9$~kpc.
Since this part of the gas disk is stripped first, we expect it to be located in the northern half of the tail at the time of interest.
If the initial model profile had a constant gas surface density of $10$~M$_{\odot}$pc$^{2}$ up to the optical radius as, e.g. NGC~3521 (Leroy et al. 2008),
we expect to observe higher H{\sc i} surface densities in the northern half of the model gas tail.

No instabilities which give rise to turbulent motions of the stripped gas are included in the model. Once the model gas clouds are pushed
out of the galactic disk, they react like ballistic particles of constant surface density subjected to the ram pressure wind. The resemblance between the observed
and the modeled gas plume thus seems surprising, because turbulence might be expected in the galaxy's wake which should lead to rapid mixing
of the ISM with the ICM leading to a lower cloud surface density $\Sigma_{\rm cl}$. Since the acceleration by ram pressure is $a_{\rm rp}=p_{\rm ram}/\Sigma_{\rm cl}$,
the mixed clouds are expected to be stripped more efficiently. This would lead to a larger extension of the gas tail at the time of interest and
an earlier time of interest might have to be chosen based on the extension of the H{\sc i} tail which could
potentially lead to a contradiction with the observed star formation quenching age.

Recent studies of the X-ray emission of the leading edges of ram pressure stripped elliptical galaxies showed
that the viscosity is at most $5$ to $10$\,\% of the isotropic Spitzer-like viscosity (Roediger et al. 2015, Su et al. 2017, Ichinoh et al. 2017).
In addition, thermal conduction is suppressed by at least a factor of $\sim 50$ (Vikhlinin et al. 2001, Sun et al. 2005, 2007)
and X-ray tails of stripped gas do not become hotter with the increasing distance to the galaxy (Sun et al. 2010, Zhang et al. 2013).
Magnetic fields are most probably responsible for the suppression of viscosity and heat conduction (see, e.g., Pope et al. 2005, Vijayaraghavan \& Sarazin 2017).
In the light of these facts, the resemblance between the modeled and the observed tail can be explained by inefficient gas mixing and thermal conduction
of the stripped gas.

The total mass of the model gas plume between $7.5$~kpc$ \le z \le 100$~kpc with column densities in excess of $3 \times 10^{19}$~cm$^{-2}$ is $7.7 \times 10^8$~M$_{\odot}$,
a factor of $\sim 2$ higher than the observed H{\sc i} in the extended gas plume. The model gas mass between $7.5$~kpc$ \le z \le 100$~kpc is
in agreement with the observed gas mass for a limiting column density of $8 \times 10^{19}$~cm$^{-2}$. This means that ISM--ICM mixing and heating through
thermal conduction does occur, but these are slow processes. Since the gas left the galactic disk about
$200$~Myr ago (the star formation quenching timescale), the mixing and thermal conduction timescales should be of the same order to decrease the atomic gas mass
by a factor of $2$.

Vollmer et al. (2001) estimated the timescale of saturated evaporation by the intracluster medium $t_{\rm evap}^{\rm sat} \sim 10 \times (N/(10^{20}~{\rm cm}^{-2}))$~Myr,
where $N$ is the column density of the stripped gas based on the results of Cowie \& McKee (1977). 
If the H{\sc i} clouds were stripped with a typical column density of $10^{21}$~cm$^{-2}$ within galactic disks, 
the evaporation time is $t_{\rm evap}^{\rm sat} \sim 100$~Myr.
As already stated above, in the presence of a magnetic field configuration that inhibits
heat flux (e.g., a tangled magnetic field), this evaporation time can increase significantly (Cowie, McKee, \& Ostriker 1981).
We suggest that the observed presence of H{\sc i} at vertical distances of $\sim 70$~kpc is evidence for an evaporation timescale
that is comparable or somewhat larger than the star formation quenching timescale of $200$~Myr. 
With a $25$ to $50$ times suppressed thermal conduction, an evaporation time of $200$~Myr corresponds to a cloud column density of $\sim 8$ to $4 \times 10^{19}$~cm$^{-2}$,
which is close to the limiting H{\sc i} column density for consistency between the observed and model gas mass in the extended tail of NGC~4388.
We thus suggest that gas of lower column density has been evaporated by the hot intracluster medium.
We conclude that the presence of the huge low column density H{\sc i} tail of NGC~4388 with a
gas stripping timescale of $\sim 200$~Myr is consistent with a suppression of the thermal conductivity by about a factor $25$ to $50$ (see also Tonnesen et al. 2011).

\subsection{Ram pressure stripping by the ICM of M~86 is improbable}

In the present scenario the gas of NGC~4388 has been stripped by the ICM of M~87. The proximity of NGC~4388 to M~86 might 
suggest that an alternative scenario is ram pressure stripping by the ICM of M~86. In the following we give our reasons why we
think that this alternative scenario is highly improbable:
\begin{itemize}
\item
since the ICM distributions are very much peaked on M87 and M86, NGC~4388 would had to come much closer to M~86 than to M~87 
(despite the 2 times higher velocity that is equivalent to an increase of ICM density of a factor of $4$). The ram pressure time profile would be much narrower 
and there would be less stripping (less amount of encountered ICM, see Jachym et al. 2009). This is a more probabilistic argument.
\item
the trajectory given by the model wind direction is consistent with stripping by the M~87 ICM (Vollmer et al. 2009). 
\item
the H{\sc i} tail of NGC~4388 is located in front of M~86 (dust absorption). Given the radial velocities (NGC~4388: $2585$~km/s$^{-1}$ ; M~86: $-224$~km/s$^{-1}$ ), 
NGC~4388 must have passed M~86 to the south. 
The spatial extent of the H{\sc i} tail ($\sim 100$~kpc) and a timescale of $250$~Myr implies a velocity in the plane of the sky
of $\sim 380$~km/s$^{-1}$. With a radial velocity of $2585+224$~km/s$^{-1}$=$2800$~km/s$^{-1}$, the axis ratio of the tail would be $\sqrt{(x^2+y^2)/z} \sim 0.15$, 
i.e. its extent in the radial direction would be 670 kpc.
The passage of NGC~4388 south of M~86 is in contradiction to the northern tip of the H{\sc i} tail that extends north of M~86, except if stripping began much
earlier than the time of peak ram pressure.
Since the stripped H{\sc i} is never accelerated to the ICM velocity and there is always a time delay between maximum ram pressure and
effective gas stripping, the northern tip of the H{\sc i} tail should have moved even further to the south of M~86.
In the present simulations the ratio between the size of the tail and the distance travelled since closest approach is $\sim 1/2$.
\end{itemize}

\section{Conclusions}

The spectacular extended H{\sc i} plume of spiral galaxy NGC~4388 is unique in the Virgo cluster (Oosterloo \& van Gorkom 2005). 
This galaxy fulfills all criteria for having undergone a recent ram pressure stripping event caused by its highly eccentric orbit
within the Virgo cluster which led it close to the cluster center where the density of the intracluster medium is high (see, e.g., Vollmer et al. 2009):
\begin{itemize}
\item
a strongly truncated H{\sc i} (Chung et al. 2009) and H$\alpha$ disk (Yoshida et al. 2002),
\item
an asymmetric ridge of polarized radio continuum emission (Vollmer et al. 2007, Damas-Segovia et al. 2016),
\item
extended extraplanar gas toward the opposite side of the ridge of polarized radio continuum emission (Oosterloo \& van Gorkom 2005), and
\item
a recent (a few $100$~Myr) quenching of the star formation activity in the outer, gas-free galactic disk (Pappalardo et al. 2010).
\end{itemize}
Nehlig et al. (2016) showed for another Virgo cluster galaxy, NGC~4501, that the location of galactic spiral arms during the interaction between 
the interstellar medium and the intracluster medium significantly influences the evolution of the gas surface density distribution.   

We made numerical simulations of the ram pressure stripping event to investigate the influence of galactic structure on the
observed properties of NGC~4388. In each simulation the spiral arms are located differently with respect to the direction
of the galaxy's motion within the intracluster medium. 
We tested different orbits and wind angles, the angle between the ram pressure wind and the galactic disk.
The multiwavelength observations are ``best'' reproduces by a wind angle of $30^{\circ}$.

Out of ten simulations only two simulations could approximately reproduce the following
observational constraints:
\begin{itemize}
\item
the extent of the extended H{\sc i} plume and the surface density distribution of its southern half,
\item
the east-west asymmetry of the H{\sc i} disk,  
\item
the asymmetry of the H{\sc i} position-velocity diagram along the major axis of the galactic disk,
\item
the FUV emission distribution,
\item
the southern asymmetric ridge of polarized radio continuum emission, and
\item
the optical spectrum of the outer, gas-free region of the galactic disk.
\end{itemize}
Given the degeneracies between galactic structure, time of peak ram pressure, wind angle, and temporal ram pressure profile on the
outcome of the simulations, we could not find one single simulation and projection that fulfilled all our criteria.
Models~9 and 10 (Table~\ref{tab:results}) with an inclination angle of $i=75^{\circ}$ are consistent with a maximum of
the observational constraints.

These ``best-fit'' models do not reproduce the morphology of the northern half of the H{\sc i} tail, the protruding H{\sc i} feature in the
northwestern part of the galactic disk, the smaller velocity width of the far tail, and the detailed H$\alpha$ and radio continuum emission distributions.  

The derived time since the quenching of the star formation activity of the outer, gas-free disk is $\sim 200$~Myr. This time is somewhat smaller than the time
since maximum ram pressure ($\sim 240$~Myr). The peak ram pressure was $p_{\rm max}=8.3 \times 10^{-11}$~dyn\,cm$^{-2}$, the
present ram pressure is $p_{\rm today}=1.25 \times 10^{-11}$~dyn\,cm$^{-2}$. The 3D velocity vector of NGC~4388 is $\vec{v}_{\rm gal}=(0.11,-0.69,0.72)$.
The galaxy velocity within the intracluster medium is $v_{\rm gal} \sim 2200$~km\,s$^{-1}$. 

Based on our results we conclude that
\begin{enumerate}
\item
the combination of a deep optical spectrum of the outer gas-free region of the galactic disk together with deep H{\sc i},
H$\alpha$, FUV, and polarized radio continuum data permits to constrain numerical simulations to derive the temporal ram pressure profile, the 3D
velocity vector of the galaxy, and the time since peak ram pressure with a high level of confidence,
\item
for the detailed reproduction of multi-wavelength
observations of a spiral galaxy that undergoes or underwent a ram pressure stripping event,
galactic structure has to be taken into account,
\item
the observed asymmetries in the disk of NGC~4388 are not caused by
the present action of ram pressure, but by the resettling of gas that has been pushed out of the galactic disk during
the ram pressure stripping event (see also Damas-Segovia et al. 2016),
\item
the presence of the huge low column density H{\sc i} tail of NGC~4388 with a
gas stripping timescale of $\sim 200$~Myr is consistent with a suppression of the thermal conductivity by about a factor $25$ to $50$.
\end{enumerate}

\begin{acknowledgements}
BV would like to thank A. Damas for sharing the JVLA image of NGC~4388 and Ming Sun for sharing his insights on gas viscosity and thermal conduction. 
MS acknowledges the support by the Polish National Science Centre (NCN) through the grant 2012/07/B/ST9/04404.
CP acknowledges support from the Science and Technology Foundation (FCT, Portugal) through the Postdoctoral Fellowship SFRH/BPD/90559/2012,
PEst-OE/FIS/UI2751/2014, PTDC/FIS-AST/2194/2012,  and  through  the  support to the IA activity via the UID/FIS/04434/2013 fund.
\end{acknowledgements}

\clearpage

\begin{appendix}

\section{Atomic hydrogen}

\begin{figure*}
        \resizebox{\hsize}{!}{\includegraphics{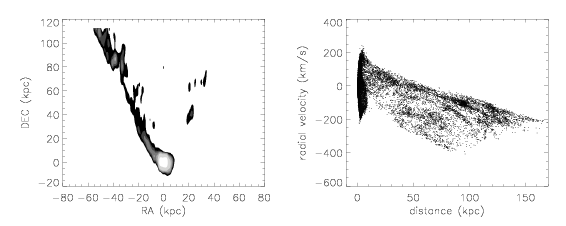}\includegraphics{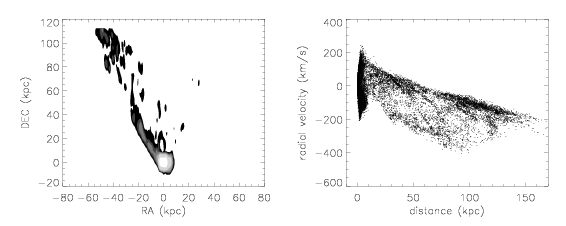}
        \put (-500,50) {\Large model 2}\put (-1060,50) {\Large model 1}}
        \resizebox{\hsize}{!}{\includegraphics{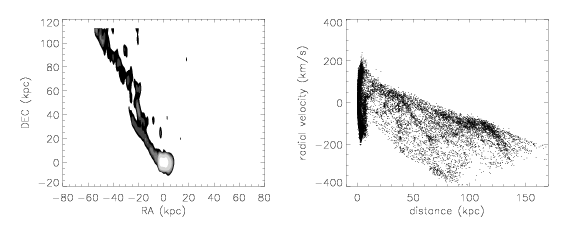}\includegraphics{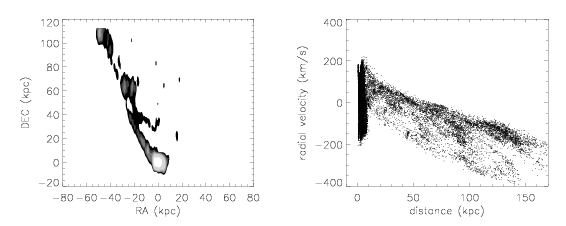}
        \put (-500,50) {\Large model 4}\put (-1060,50) {\Large model 3}}
        \resizebox{\hsize}{!}{\includegraphics{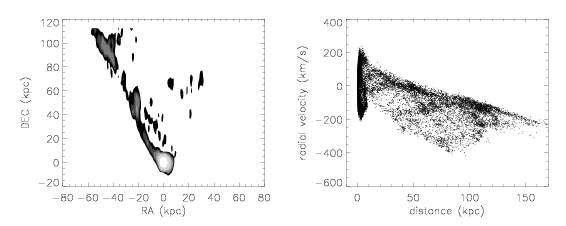}\includegraphics{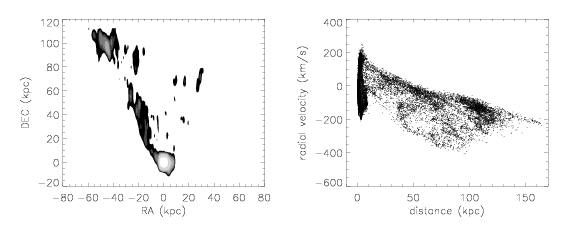}
        \put (-500,50) {\Large model 6}\put (-1060,50) {\Large model 5}}
        \resizebox{\hsize}{!}{\includegraphics{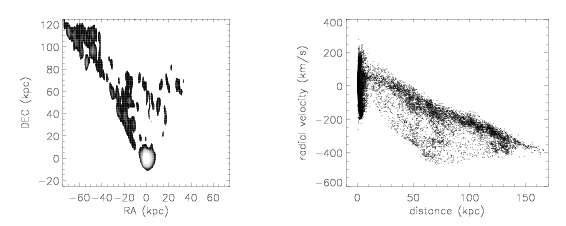}\includegraphics{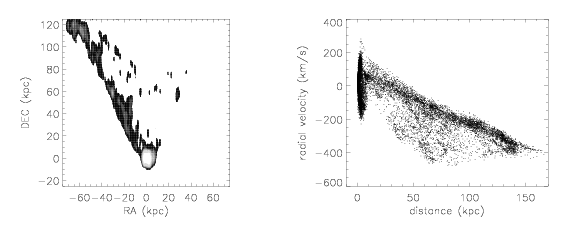}
        \put (-500,50) {\Large model 8}\put (-1060,50) {\Large model 7}}
        \resizebox{\hsize}{!}{\includegraphics{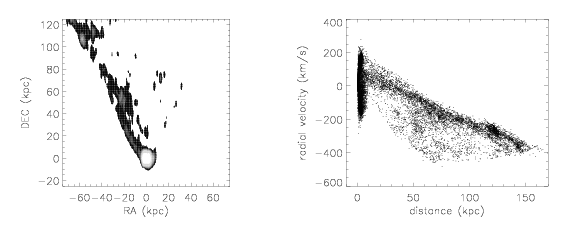}\includegraphics{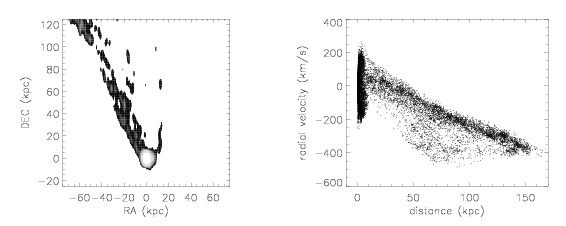}
        \put (-500,50) {\Large model 10}\put (-1060,50) {\Large model 9}}
        \caption{Model H{\sc i} large-scale distributions with $i=75^{\circ}$. Greyscale levels are (4, 7, 14, 22, 30, 36, 43, 50, 58, 65, 72, 80)$\times 10^{20}$~cm$^{-2}$.
        } \label{fig:N4388_final_test3_HI1b}
\end{figure*}

\section{H$\alpha$}

\begin{figure*}
        \resizebox{\hsize}{!}{\includegraphics{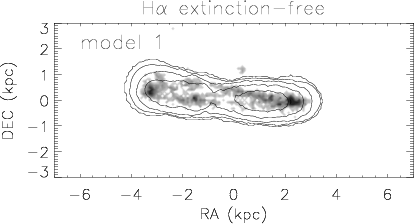}\includegraphics{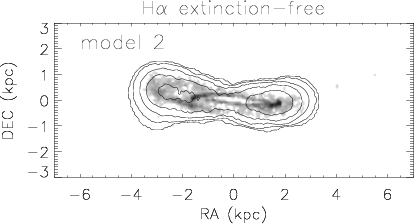}}
        \resizebox{\hsize}{!}{\includegraphics{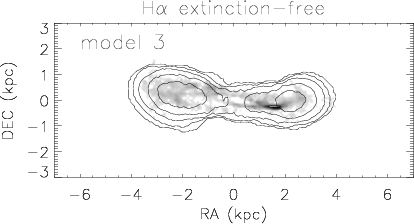}\includegraphics{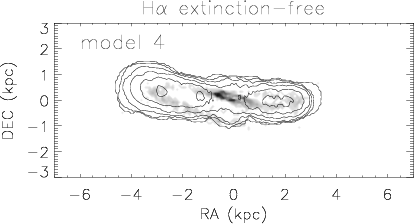}}
        \resizebox{\hsize}{!}{\includegraphics{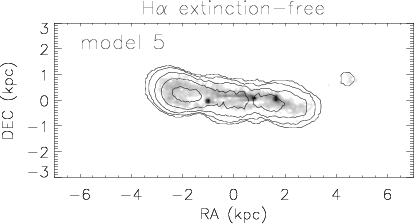}\includegraphics{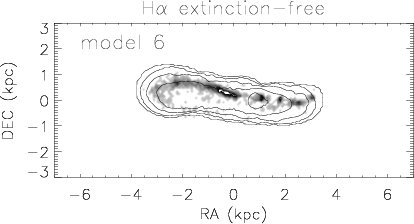}}
        \resizebox{\hsize}{!}{\includegraphics{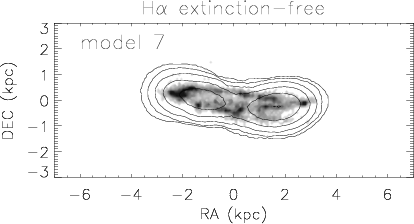}\includegraphics{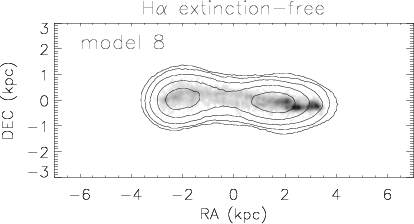}}
        \resizebox{\hsize}{!}{\includegraphics{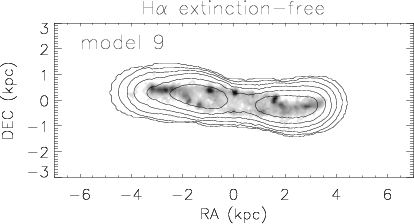}\includegraphics{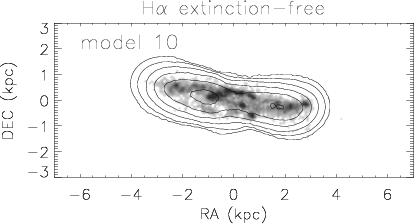}}
        \caption{Model H{\sc i} contours on model H$\alpha$ emission distributions without extinction with $i=90^{\circ}$.
        } \label{fig:N4388_final_test3_deltat1_plots_ha1}
\end{figure*} 

\section{FUV}

\begin{figure*}
        \resizebox{\hsize}{!}{\includegraphics{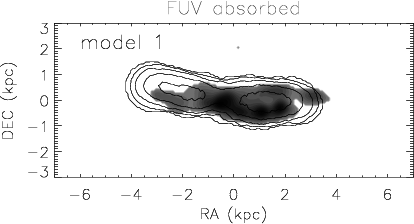}\includegraphics{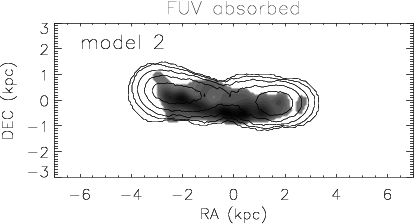}}
        \resizebox{\hsize}{!}{\includegraphics{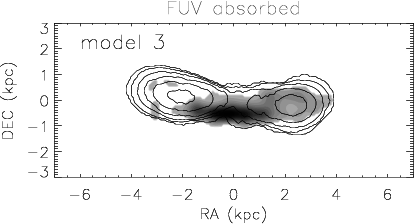}\includegraphics{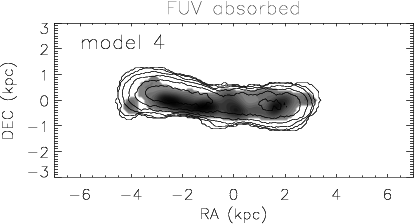}}
        \resizebox{\hsize}{!}{\includegraphics{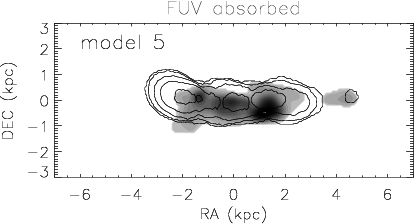}\includegraphics{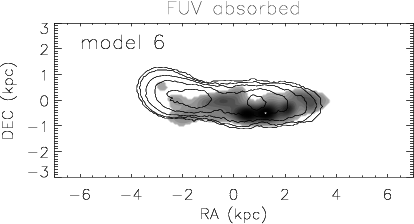}}
        \resizebox{\hsize}{!}{\includegraphics{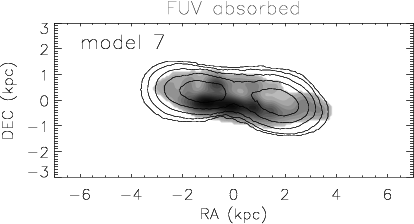}\includegraphics{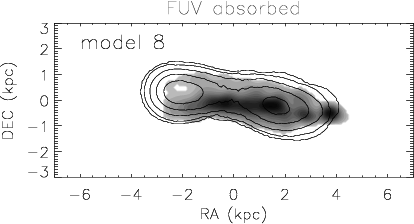}}
        \resizebox{\hsize}{!}{\includegraphics{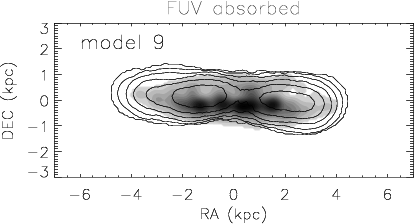}\includegraphics{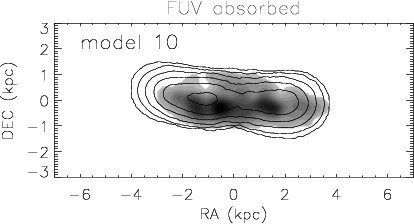}}
        \caption{Model H{\sc i} contours on model FUV emission distributions with $i=75^{\circ}$.
        } \label{fig:N4388_final_test3_deltat1_plots_fuv1}
\end{figure*}

\section{Polarized radio continuum}

\begin{figure*}
        \resizebox{16cm}{!}{\includegraphics{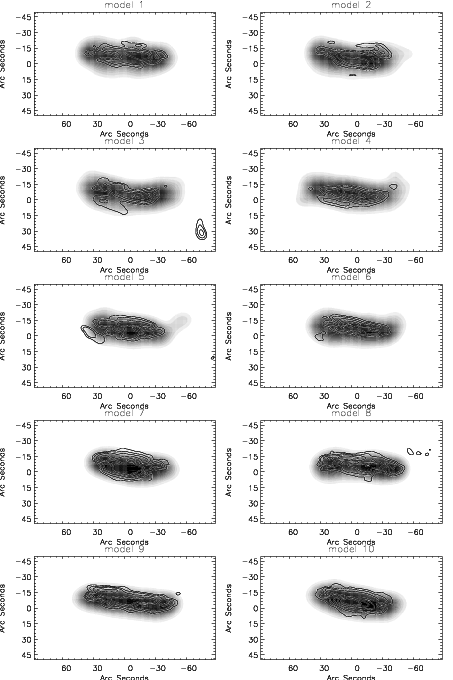}}
        \caption{Model polarized emission distributions on model H{\sc i} distribution with $i=90^{\circ}$.
        } \label{fig:pivergleich1a}
\end{figure*}

\end{appendix}

\end{document}